\documentclass[useAMS,usenatbib]{mn2e}
\usepackage[dvips]{graphicx}
\usepackage{float}
\usepackage{amsmath}	
\usepackage{amssymb}	
\usepackage{multicol}   
\usepackage{bm}		    
\usepackage{pdflscape}	
\usepackage{array}
\usepackage{booktabs}
\usepackage{tabularx}
\usepackage{natbib}
\usepackage{rotating}
\title[Monthly Notices: \LaTeXe\ guide for authors]
  {Identification of BASS DR3 Sources as Stars, Galaxies and Quasars by XGBoost}
\author[C. Li et~al.]
  {Changhua Li$^{1,2,3}$,
  Yanxia Zhang$^{1}$\thanks{Email: zyx@bao.ac.cn},
  Chenzhou Cui$^{1,3}$\thanks{Email: ccz@bao.ac.cn}, Dongwei Fan$^{1,3}$, \and Yongheng Zhao$^{1}$, Xue-Bing Wu$^{4,5}$, Boliang He$^{1,2,3}$, Yunfei Xu$^{1,3}$, Shanshan Li$^{1,2,3}$, \and Jun Han$^{1,3}$, Yihan Tao$^{1,3}$, Linying Mi$^{1,3}$, Hanxi Yang$^{1,3}$, Sisi Yang$^{1,3}$ \\
  $^1$ National Astronomical Observatories, Beijing, 100101, China\\
  $^2$ University of Chinese Academy of Sciences, Beijing 100049, China\\
  $^3$ National Astronomical Data Center, Beijing 100101, China\\
  $^4$ Department of Astronomy, School of Physics, Peking University, Beijing 100871, China\\
  $^5$ Kavli Institute for Astronomy and Astrophysics, Peking University, Beijing 100871, China}
\date{Released 2002 Xxxxx XX}

\begin{document}

\label{firstpage}

\maketitle

\begin{abstract}
The Beijing-Arizona Sky Survey (BASS) Data Release 3 (DR3) catalogue was released in 2019, which contains the data from all BASS and the Mosaic $z$-band Legacy Survey (MzLS) observations during 2015 January and 2019 March, about 200 million sources. We cross-match BASS DR3 with spectral databases from the Sloan Digital Sky Survey (SDSS) and the Large Sky Area Multi-object Fiber Spectroscopic Telescope (LAMOST) to obtain the spectroscopic classes of known samples. Then, the samples are cross-matched with ALLWISE database. Based on optical and infrared information of the samples, we use the XGBoost algorithm to construct different classifiers, including binary classification and multiclass classification. The accuracy of these classifiers with the best input pattern is larger than 90.0 per cent. Finally, all selected sources in the BASS DR3 catalogue are classified by these classifiers. The classification label and probabilities for individual sources are assigned by different classifiers. When the predicted results by binary classification are the same as multiclass classification with optical and infrared information, the number of star, galaxy and quasar candidates is separately 12 375 838 ($P_{\rm S}>0.95$), 18 606 073 ($P_{\rm G}>0.95$) and 798 928 ($P_{\rm Q}>0.95$). For these sources without infrared information, the predicted results can be as a reference. Those candidates may be taken as input catalogue of LAMOST, DESI or other projects for follow up observation. The classified result will be of great help and reference for future research of the BASS DR3 sources.

\end{abstract}


\section{Introduction} \label{sec:intro}
The classification is one of the most fundamental steps in astronomical data analysis. It is helpful both for studies of individual systems and for statistical population analyses by providing classification labels for large numbers of sources. Especially the study of quasars is of great significance for answering many astronomical scientific questions, such as cosmology, understanding the physical nature of X-ray sources and variable sources. Since the first quasar was identified in the 1960s, astronomers have taken a lot of efforts to find quasars as many as possible. With the advent of the survey telescopes, a large number of quasars have been discovered in recent 20 years. The sky survey telescopes include the Large Bright Quasar Survey (LBQS; \citealt{Hewett1995, Hewett2001}), the Anglo-Australian Telescope Survey (AATS; \citealt{AATS1990}), the FIRST Bright Quasar Survey (FBQS; \citealt{FBQS1996}), the Palomar Scan Grism Surveys (PSGS; \citealt{PSGS1999}), the Sloan Digital Sky Survey (SDSS; \citealt{york00}) and Large Sky Area Multi-Object Fiber Spectroscopic Telescope (LAMOST; \citealt{Cui12}). Through the implementation of these survey projects, more and more quasar candidates have been identified. In fact, as long as there is enough observational data of celestial objects, such as spectra and multi-wavelength data, it is not a difficult task to distinguish the astronomical sources. However, it is hard and time-consuming work to obtain spectroscopic observation for a large volume of sources, especially for faint sources. It is necessary to classify the photometric data without spectra before understanding them. For these photometric data, machine learning is a good solution to classify them with the identified sources. In contrast to traditional methods, machine learning is comparatively fast and applicable.

Machine learning (ML) is a method of realizing artificial intelligence (AI), which is mainly applied to problems that are difficult to describe with rules and program explicitly. Its goal is to study how to make computers simulate human learning behaviors, automatically improve algorithms through experience, learn hidden patterns from data and build models, so as to be able to make prediction on similar problems. There are many applications in astronomy utilizing ML and AI, including the discovery of extrasolar planets \citep{Pearson2018, Shallue2018} and gravitational lens systems \citep{Jacobs2017, Lanusse2018, Pourrahmani2018}; discovery and classification of transient objects \citep{Connor2018, Farah2018, Farah2019, Mahabal2019}; forecasting solar activity \citep{Florios2018, Inceoglu2018, Nishizuka2017}; assignment of photometric redshifts within large-scale galaxy surveys \citep{Bilickiet2018, Ruiz2018, Speagle2017}; and classification of gravitational wave signals and instrumental noise \citep{George2018a, George2018b, Powell2017}. In terms of celestial object classification, there have been many research works on this respect, such as distinguishing stars and galaxies by artificial neural network (ANN; \citealt{Odewahn1994}), decision tree (DT; \citealt{Weir1995}), random forest \citep{Clarke2020} or the Kohonsen self-organizing map (SOM; \citealt{Miller1996}), separating quasars/AGNs from stars by learning vector quantization (LVQ), single-layer perception (SLP), or support vector machines (SVMs) \citep{Zhang2004}, targeting quasar candidates by support vector machine (SVM) \citep{Gao2008, Peng2012, Jinx2019} and XGBoost \citep{Jinx2019} and so on.

The Beijing-Arizona Sky Survey (BASS; \citealt{Huzou2017a}) is a wide-field two-band photometric survey of the northern Galactic Cap. The Mosaic $z$-band Legacy Survey (MzLS) covers the same area in $z$ band. The two surveys will be served as two of the three imaging surveys to provide photometric input catalogues for target selection of the Dark Energy Spectroscopic Instrument (DESI) project.

In this paper, we download BASS DR3 coadded catalogue, then cross-match it with SDSS, LAMOST and ALLWISE databases, obtain the spectroscopic classes of known sources, optical and infrared photometric information. We create different classifiers with known spectroscopic classes of samples based on only optical information, combined optical and infrared information. Finally these classifiers are applied to classify the BASS DR3 sources. Section~2 describes the data used and the distribution of various objects in 2-d spaces. Section~3 introduces the XGBoost algorithm. Section~4 compares the performance of XGBoost with different input patterns, creates classifiers with optimal input patterns, and discusses our results. Section~5 applies the created classifiers to the unidentified sources of BASS DR3 sources. Section~6 summarizes our work.

\section{The data} \label{sec:data}
The Beijing-Arizona Sky Survey (BASS; \citealt{Huzou2017a, Huzou2017b, Huzou2018}) used the 2.3m Bok telescope to take $g$ and $r$ band imaging over a sky area of about 5400 deg$^2$ in the northern Galactic cap at ${\delta} >$ 30$^{\circ}$, MzLS used the 4 m Mayall telescope to obtain $z$ band imaging over a similar sky area to BASS (${\delta} >$ 32$^{\circ}$). The BASS observations were carried out in the first semester of each year, from 2015 January through 2018 June. In 2017, the first data release (DR1) was released, which contains calibrated images obtained in 2015 and 2016, and their corresponding single-epoch and coadded catalogues. After a year, the second data release (DR2) was released, which includes stacked images, coadded catalogs, and single-epoch images and catalogues. In 2019, the BASS DR3 was released, which contains the data from all BASS and MzLS observations during 2015 January and 2019 March. The DR3 includes single-epoch photometric catalogue and co-added photometric catalogue. Sources in DR3 are detected in stacked images and are required to be identified in at least two bands.

The Sloan Digital Sky Survey (SDSS; \citealt{york00}) has been a most successful photometric and spectroscopic sky survey ever made, providing deep multi-colour images of one third of the sky and spectra for more than three million celestial objects. The SDSS began regular survey operations in 2000, which has progressed through several phases, SDSS-I (2000-2005), SDSS-II (2005-2008), SDSS-III (2008-2014), and SDSS-IV (2014-2020). SDSS Data Release 16 (DR16) contains about 880 652 stars, 2 616 381 galaxies and 749 775 quasars when $zWarning=0$ in DR16 SpecObj database \citep{Blan17}. The DR16 quasar catalogue (DR16Q) includes 750 414 quasars, among which 225 082 are new discoveries \citep{Brad2020}.

The Large Sky Area Multi-object Fiber Spectroscopic Telescope (LAMOST; \citealt{Cui12,Luo15}) may observe 4000 spectra in single observation to a limiting magnitude as faint as $r=19$ at the resolution $R=1800$. The first phase sky survey has been finished in five years. LAMOST survey contains the LAMOST ExtraGalactic Survey (LEGAS) and the LAMOST Experiment for Galactic Understanding and Exploration (LEGUE) survey of Milky Way stellar structure. The fifth data release (DR5) was published online in 2017 (http://dr5.lamost.org/). DR5 includes 8 183 160 stars (7 146 482 stars with S/N$>10$ in $g$ band or $i$ band), 152 863 galaxies, 52 453 quasars, and 637 889 unknown objects.

The Wide-field Infrared Survey Explorer (WISE; \citealt{Wright10}) is an entire mid-infrared sky survey, which obtained over a million images and observed hundreds of millions of celestial objects. The WISE survey provides mid-infrared information about the Solar System, the Milky Way, and the Universe. Based on the WISE work, the AllWISE program has created new products with better photometric sensitivity and accuracy as well as better astrometric precision than WISE.

We use co-added photometric catalogue from BASS DR3, which contains about 200 millions sources \citep{HuZou2019}. According to the median 5$\sigma$ AB magnitude depths \citep{HuZou2019}, we handle the BASS DR3 catalogue by removing out-of-range or bad pixel data as follows: $0<gPSFMag \le 24.2$, $0<rPSFMag \le 23.6$, $0<zPSFMag \le 23$, Flag\_ISO\_g $=0$ (Flag for Isophotal magnitude in $g$ band), Flag\_Model\_g $=0$ (Flag for PSF magnitude in $g$ band), Flag\_ISO\_r $=0$ (Flag for Isophotal magnitude in $r$ band), Flag\_Model\_r $=0$ (Flag for PSF magnitude in $r$ band), Flag\_ISO\_z $=0$ (Flag for Isophotal magnitude in $z$ band), and Flag\_Model\_z $=0$ (Flag for PSF magnitude in $z$ band). Thus the number of selected sources of BASS DR3 is 110 896 598 for classification.

According to the region of survey, BASS DR3 has intersection with known SDSS and LAMOST samples in north galactic region, respectively.
The known quasar sample includes quasars from SDSS DR16Q and identified quasars from LAMOST DR5, known galaxy sample is the sources spectroscopically identified as galaxies from SDSS DR16 and LAMOST DR5, and known star sample is the sources spectroscopically identified as stars from SDSS DR16 and LAMOST DR5. The websites of all used datasets are showed in Table~1.
The sources from different databases may be correlated by positional cross-match. The cross-match radius between BASS and SDSS databases is set as 2 arc because the corresponding sources in 2 arc occupy 92.0 per~cent for their positional offset, similarly the cross-match radius between BASS and ALLWISE databases is adopted as 4 arc since the corresponding sources take up 94.8 per~cent, and the nearest ones in the radius are kept. By the software TOPCAT \citep{Tay05}, BASS DR3 is cross-matched with known LAMOST and SDSS samples in 2 arcsec radius, respectively. Thus we obtain known BASS-LAMOST sample and known BASS-SDSS sample with ``CLASS" as spectral class. When the objects exist in both BASS-SDSS and BASS-LAMOST samples, the objects in BASS-SDSS sample are only kept. Through deleting the repetitive sources, we obtain BASS-SDSS-LAMOST sample, named Sample I, and then this sample is cross-matched with ALLWISE in 4 arcsec radius by CDS Upload X-Match of the software TOPCAT. We get the BASS-SDSS-LAMOST-ALLWISE sample as the known sample with identified spectral classes, named Sample II. All photometries in the samples are extinction-corrected according to the work \citep{Sch17} and AB magnitudes are adopted. Finally, all sample information are listed in Table~2. The sample columns information is shown in Table~3.

\begin{table}
\begin{center}
\caption[]{Websites for catalogues \label{tab:Web}}
 \begin{tabular}{l}
 \hline
BASS-DR3 catalogue\\
https://nadc.china-vo.org/data/data/bassdr3coadd/f\\
\hline
Known stars, galaxies and quasars from SDSS\\
http://skyserver.sdss.org/dr16/en/tools/search/sql.aspx\\
\hline
Known stars, galaxies and quasars from LAMOST\\
http://dr5.lamost.org/v3/catalogue\\
\hline
SDSS DR16 Quasar catalog (DR16Q)\\
https://www.sdss.org/dr16/algorithms/qso\_catalog\\
\hline
\end{tabular}
\end{center}
\end{table}

\begin{table*}
\begin{center}
\caption{The parameters, definition, catalogues and wavebands}
\begin{tabular}{rlll}
\hline
Parameter&Definition  &Catalogue& Waveband\\
\hline
id &Source ID      &BASS   &\\
ra &Right ascension in decimal degrees &BASS &\\
dec &Declination in decimal degrees    &BASS &\\
$gKronMag$ &Kron magnitude in $g$ band  &BASS &Optical band\\
$rKronMag$ &Kron magnitude in $r$ band  &BASS &Optical band\\
$zKronMag$ &Kron magnitude in $z$ band  &BASS &Optical band\\
$gPSFMag$   &PSF magnitude in $g$ band  &BASS &Optical band\\
$rPSFMag$   &PSF magnitude in $r$ band  &BASS &Optical band\\
$zPSFMag$   &PSF magnitude in $z$ band  &BASS  &Optical band\\
$W1mag$   &$W1$ magnitude  &ALLWISE &Infrared band\\
$W2mag$   &$W2$ magnitude  &ALLWISE &Infrared band\\
CLASS &The spectral class label  &SDSS, LAMOST &\\
$g$    & extinction-corrected PSF magnitude in $g$ band  &BASS & Optical band\\
$r$    & extinction-corrected PSF magnitude in $r$ band  &BASS & Optical band\\
$z$    & extinction-corrected PSF magnitude in $z$ band  &BASS & Optical band\\
$W1$   & extinction-corrected $W1$ magnitude &ALLWISE& Infrared band\\
$W2$   & extinction-corrected $W2$ magnitude &ALLWISE& Infrared band\\
\hline
\end{tabular}
\bigskip
\end{center}
\end{table*}

For brief, we define ${\Delta}g=gPSFMag-gKronMag$, ${\Delta}r=rPSFMag-rKronMag$, ${\Delta}z=zPSFMag-zKronMag$. For resolved sources (e.g. galaxies), the Kron magnitude is a better measure, while for unresolved point sources (e.g. stars and quasars), the PSF magnitude is the best measure by fitting a point spread function (PSF) to the sources. The difference between $PSFMag$ and $KronMag$ in the same band has great impact on the distinction between point sources and extended sources. For example, galaxies are clearly distinguished from stars and quasars by the distribution of $iPSFMag-iKronMag$ and $zPSFMag-zKronMag$ from Pan-STARRS database \citep{Jinx2019}. For our sample, Figure~1 shows statistical distribution among point sources (stars and quasars) and extended sources (galaxies). As shown in Figure~1, the difference between Kron magnitude and PSF magnitude in different bands peaks at about zero for stars and quasars, and about 0.6 for galaxies. When ${\Delta}g>0.15$ or ${\Delta}r > 0.20$, or ${\Delta}z > 0.20$, more than 91.0 per cent of sources can be identified correctly, but part of sources are still misclassified. These results are consistent with the fact that galaxies are extended and stars and quasars are pointed from morphology. Figure~2 indicates the distribution of galaxies, stars and quasars in different 2-d spaces, which indicates that galaxies, stars and quasars are not separable with any feature alone or two of them for most of data are overlapping, nevertheless, the difference of them is more clear with additional infrared features. Especially for stars and quasars, they are almost mixed together in some 2-d spaces and become easy to discriminate when adding infrared information. Luminous quasars are apparently different from stars by means of WISE colours. All of features are helpful more or less to the classification. So in this paper, we aim to use machine learning to classify them with all available information.

\begin{figure*}
\centering
\includegraphics[height=6cm,width=5.5cm]{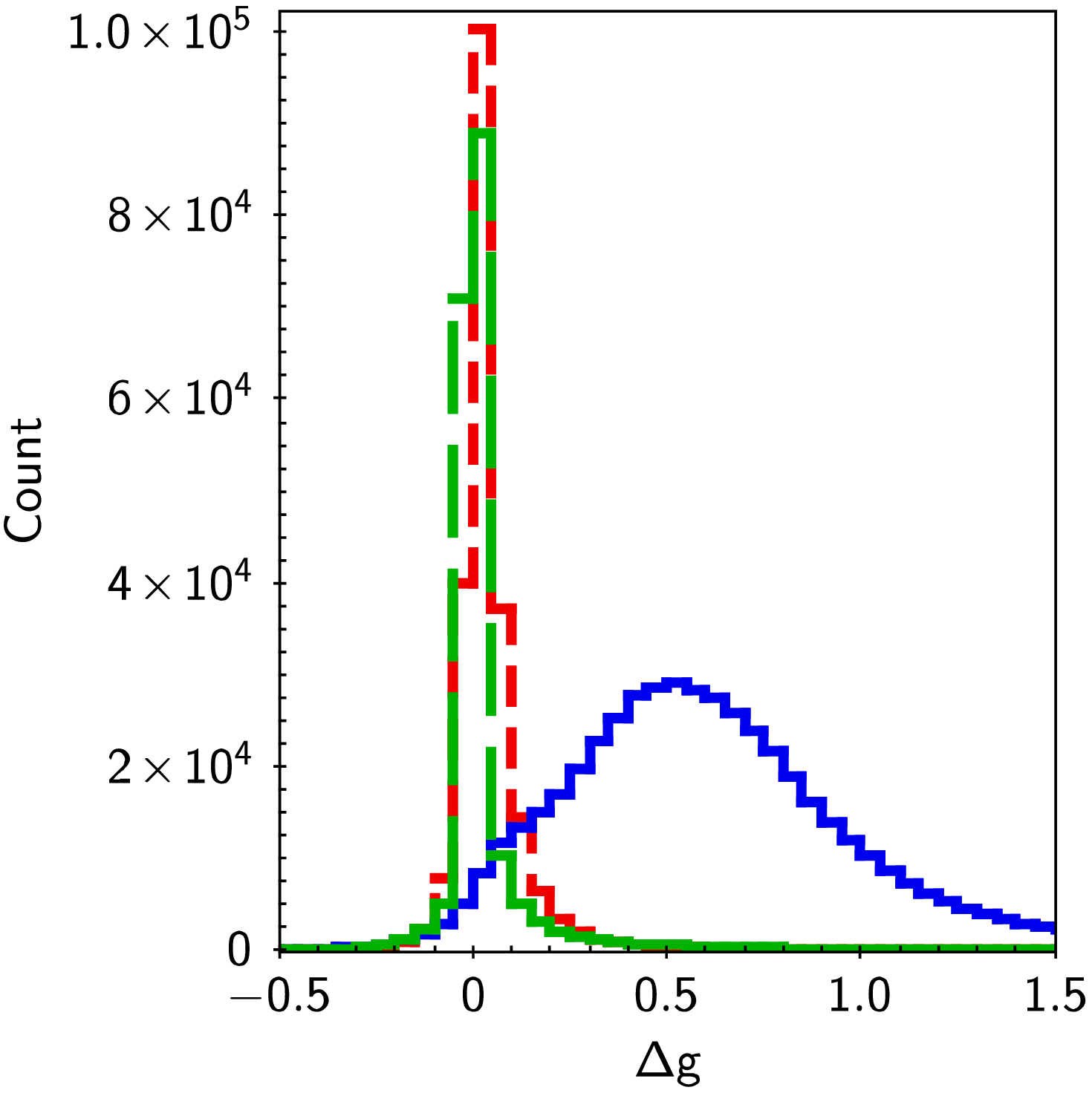}
\includegraphics[height=6cm,width=5.5cm]{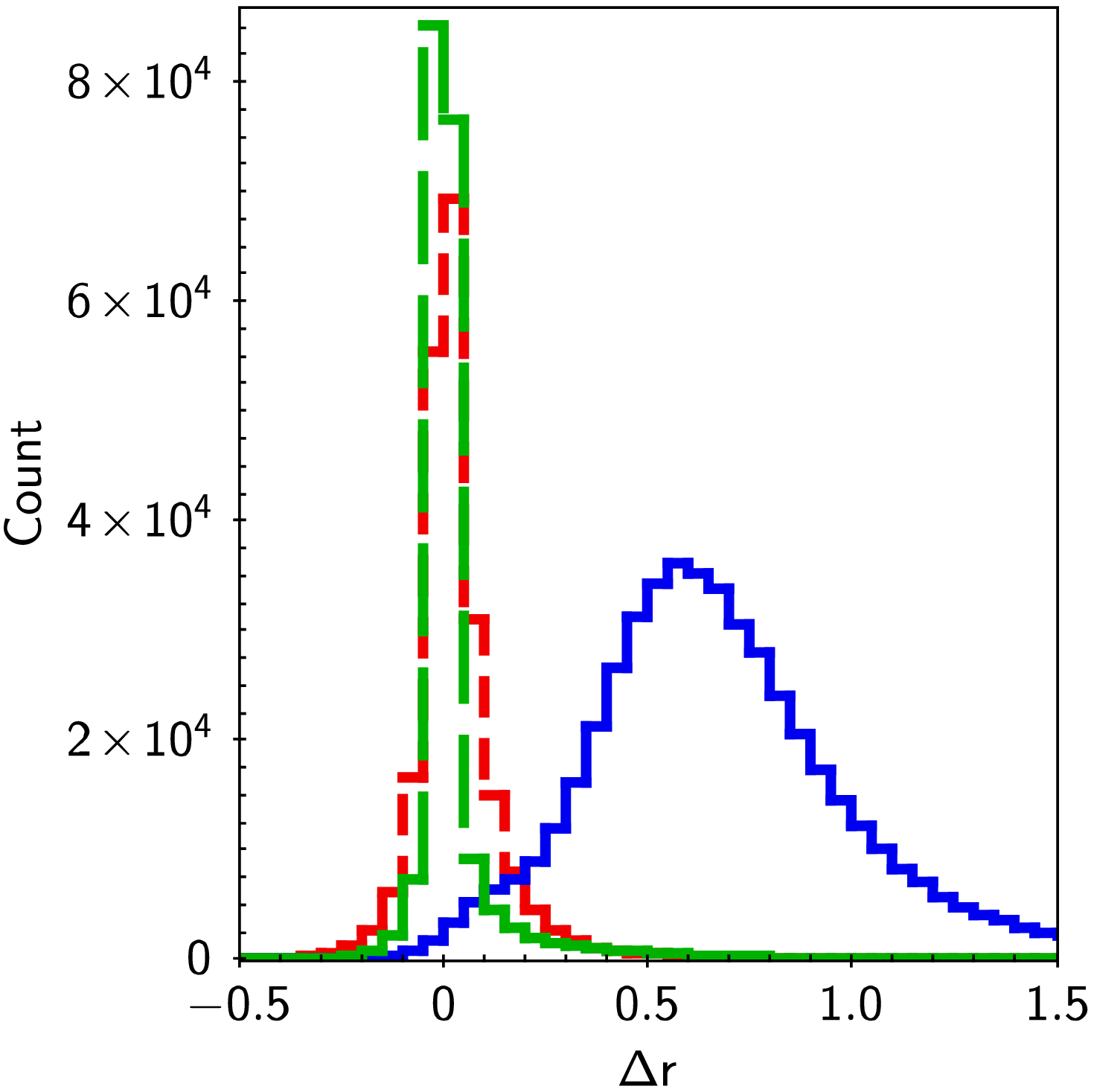}
\includegraphics[height=6cm,width=5.5cm]{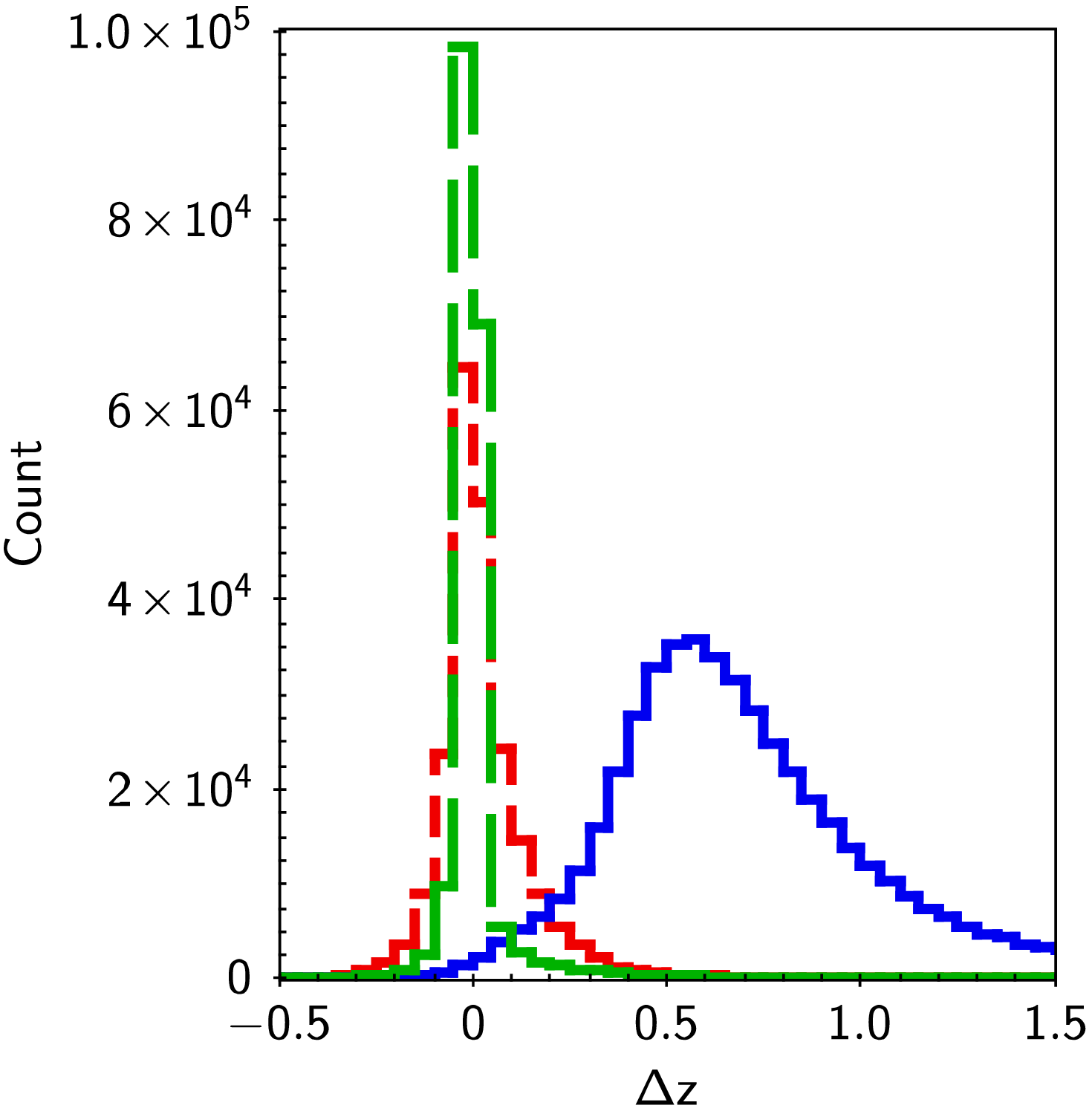}
\caption{The distribution of ${\Delta}g$, ${\Delta}r$, ${\Delta}z$ of known stars, quasars and galaxies. The green long dash line represents stars, the blue line represents galaxies and the red dash line represents quasars. }
\label{fig3}
\end{figure*}

\begin{figure*}
	\centering
	\includegraphics[height=5.5cm,width=5.5cm]{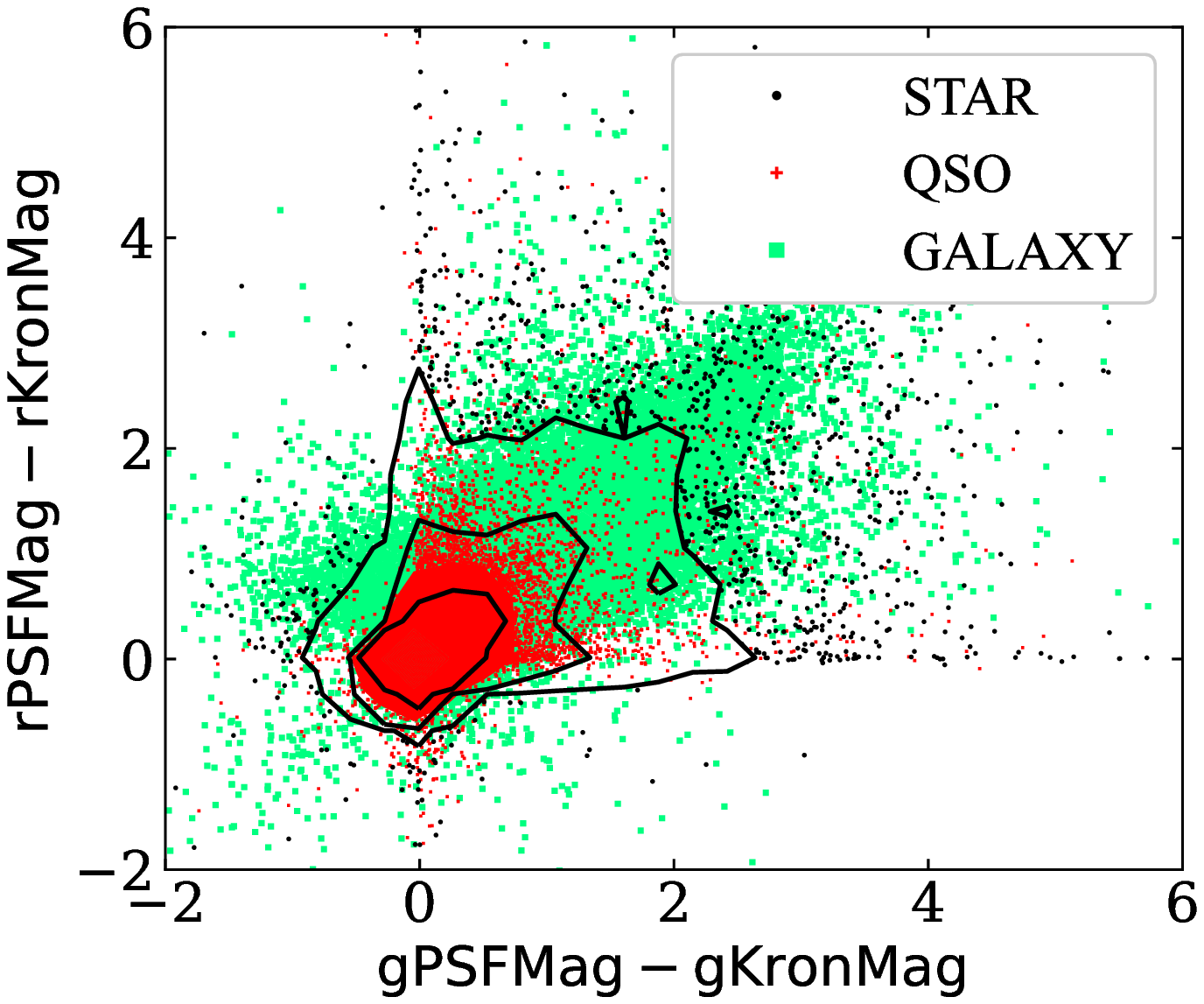}
	\includegraphics[height=5.5cm,width=5.5cm]{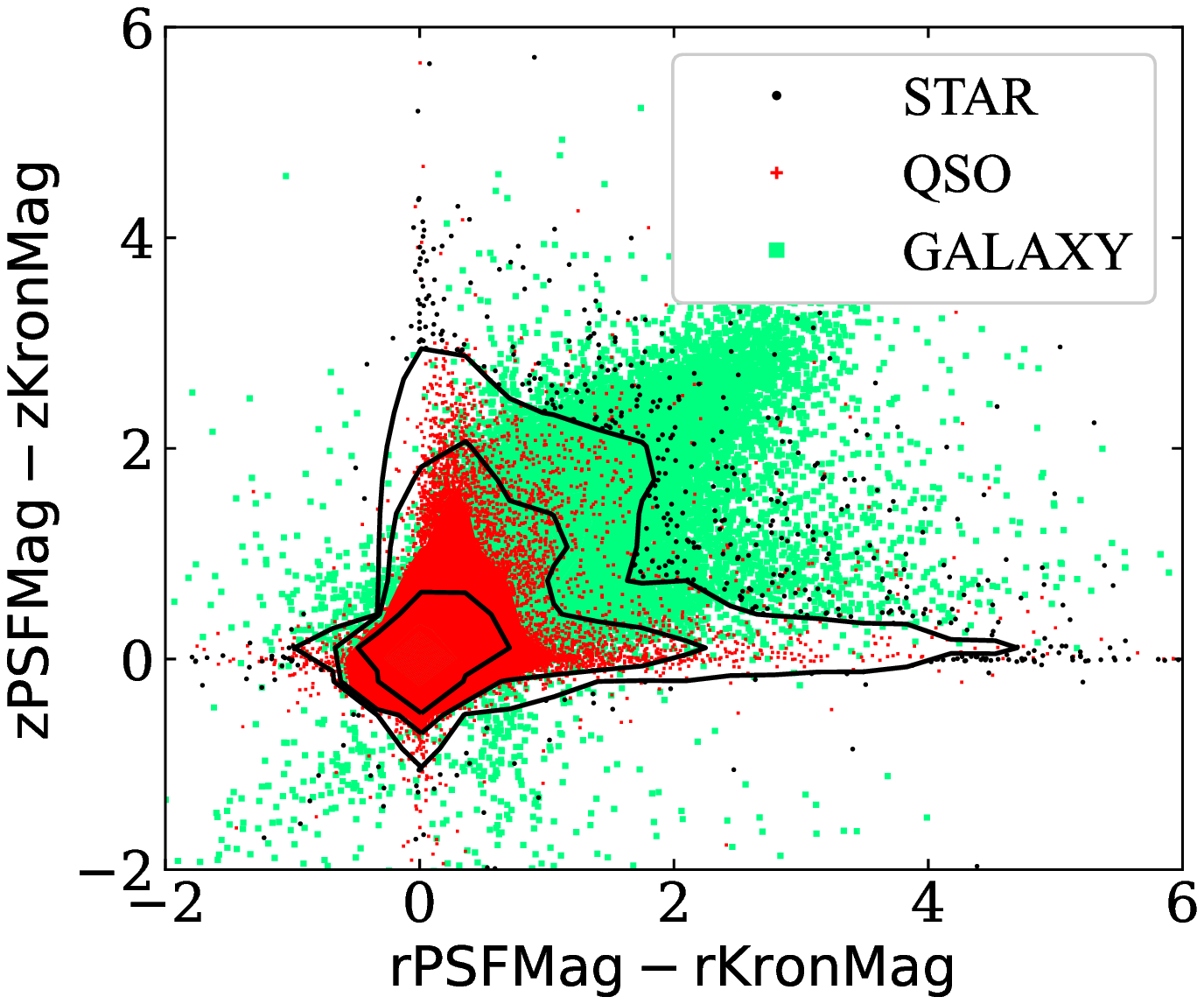}
	\includegraphics[height=5.5cm,width=5.5cm]{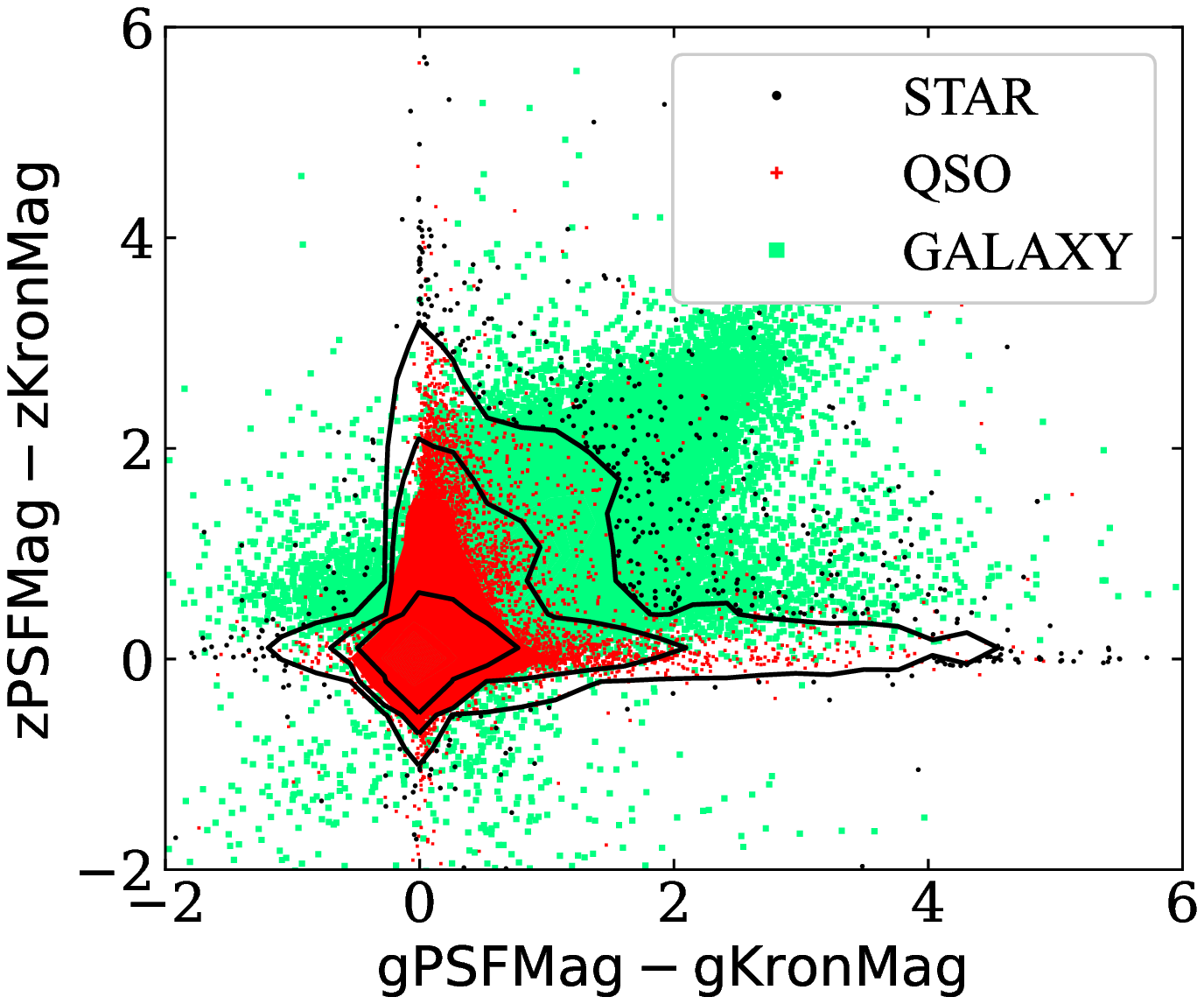}
	\includegraphics[height=5.5cm,width=5.5cm]{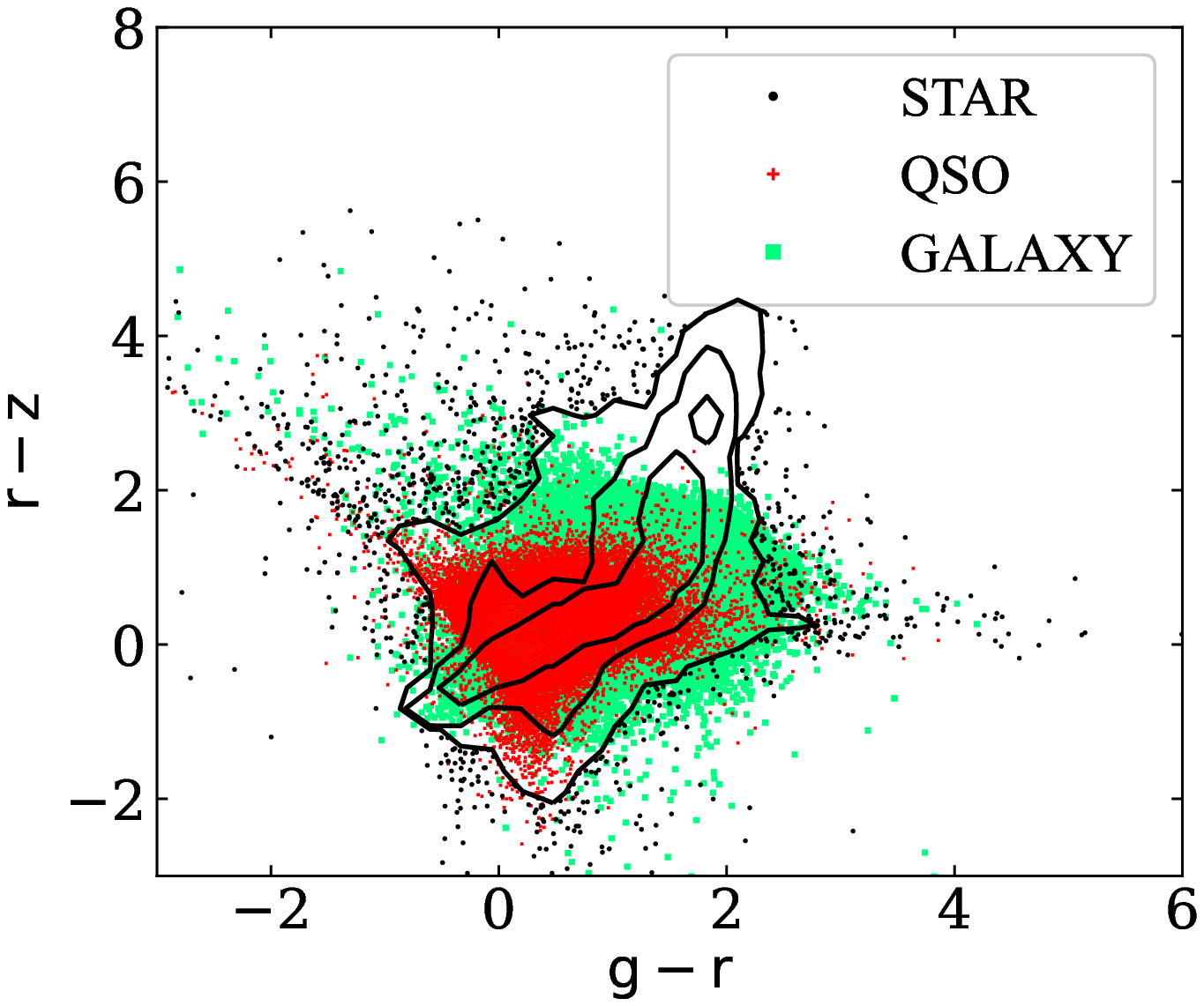}
	\includegraphics[height=5.5cm,width=5.5cm]{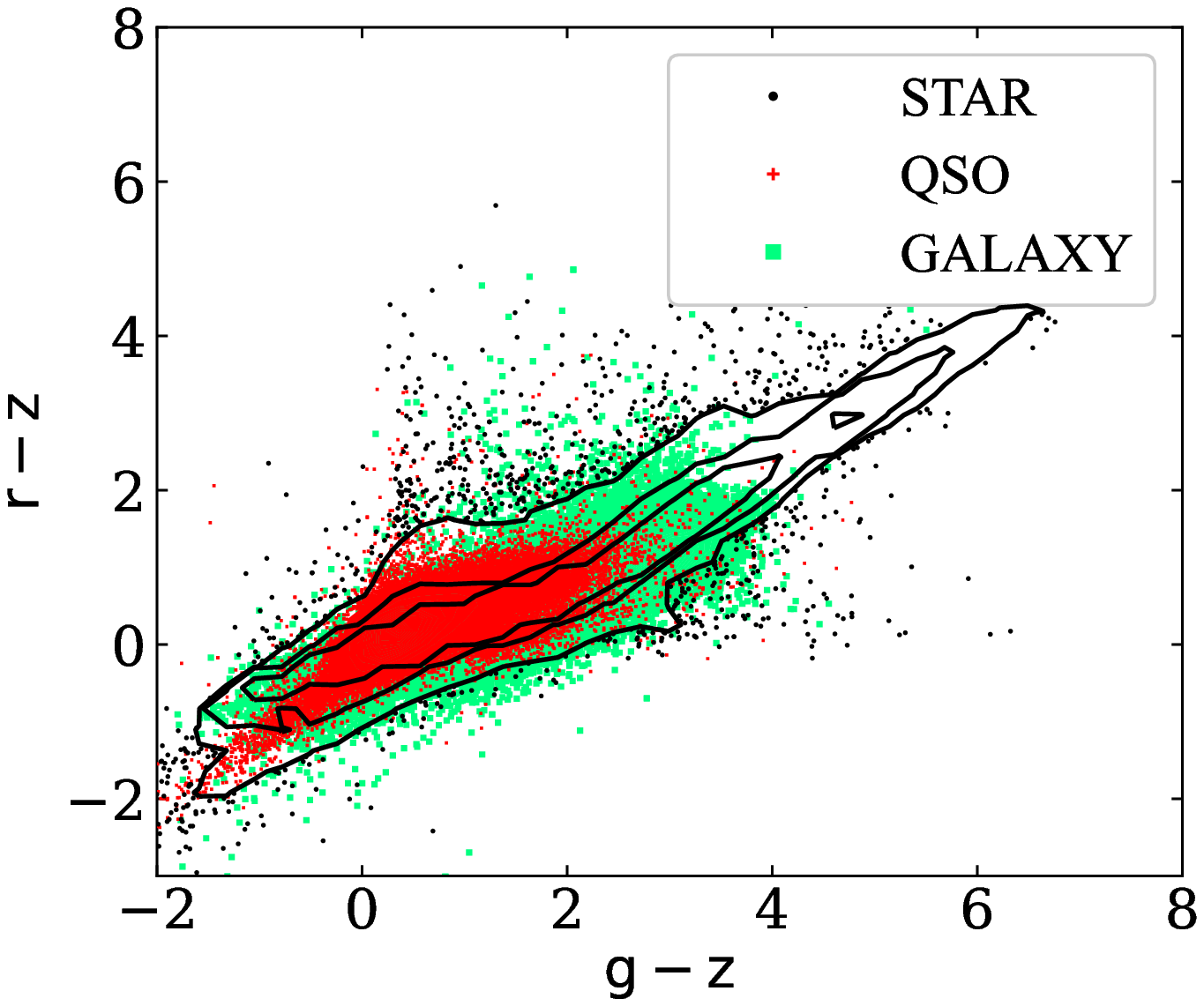}
	\includegraphics[height=5.5cm,width=5.5cm]{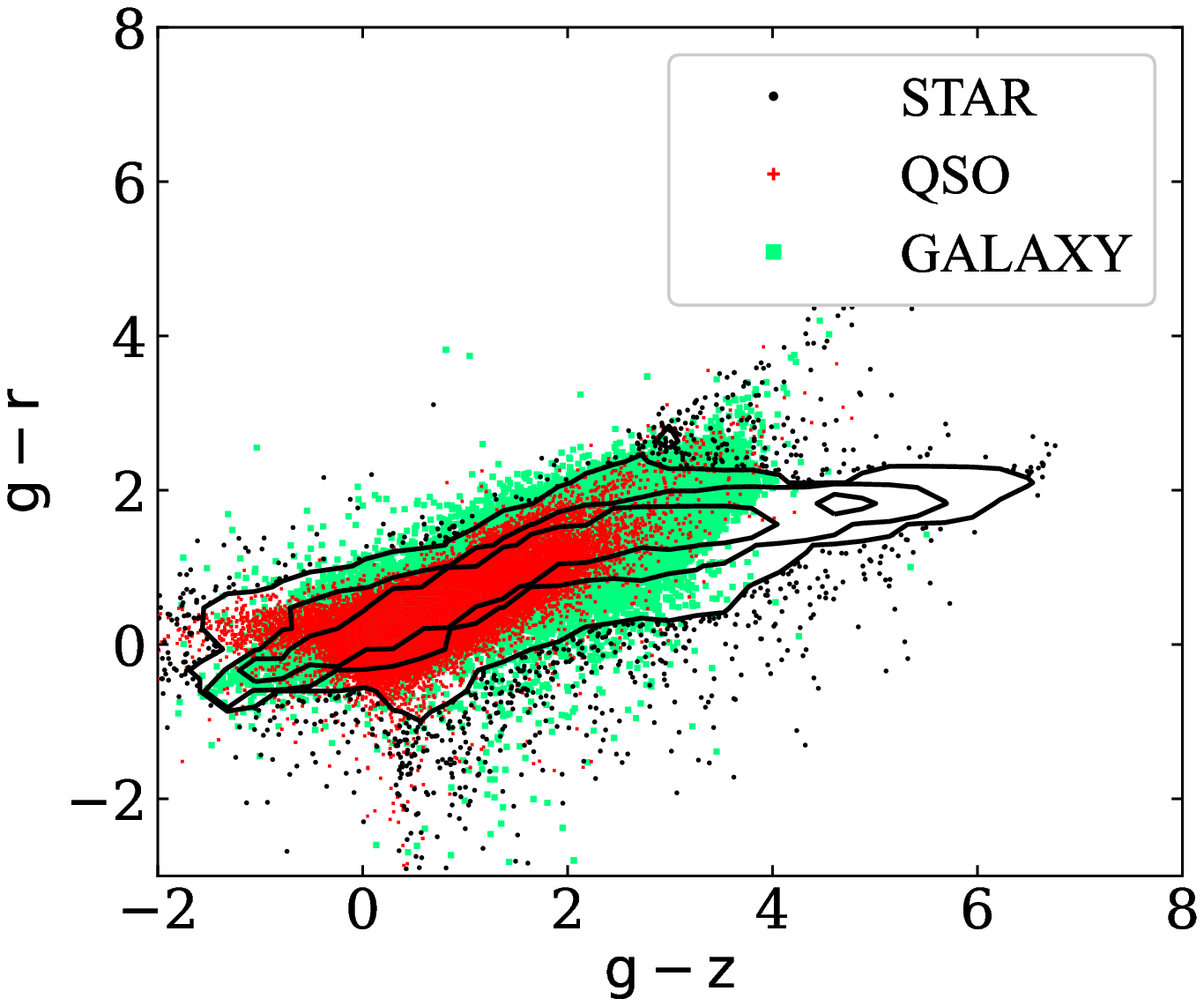}
	\includegraphics[height=5.5cm,width=5.5cm]{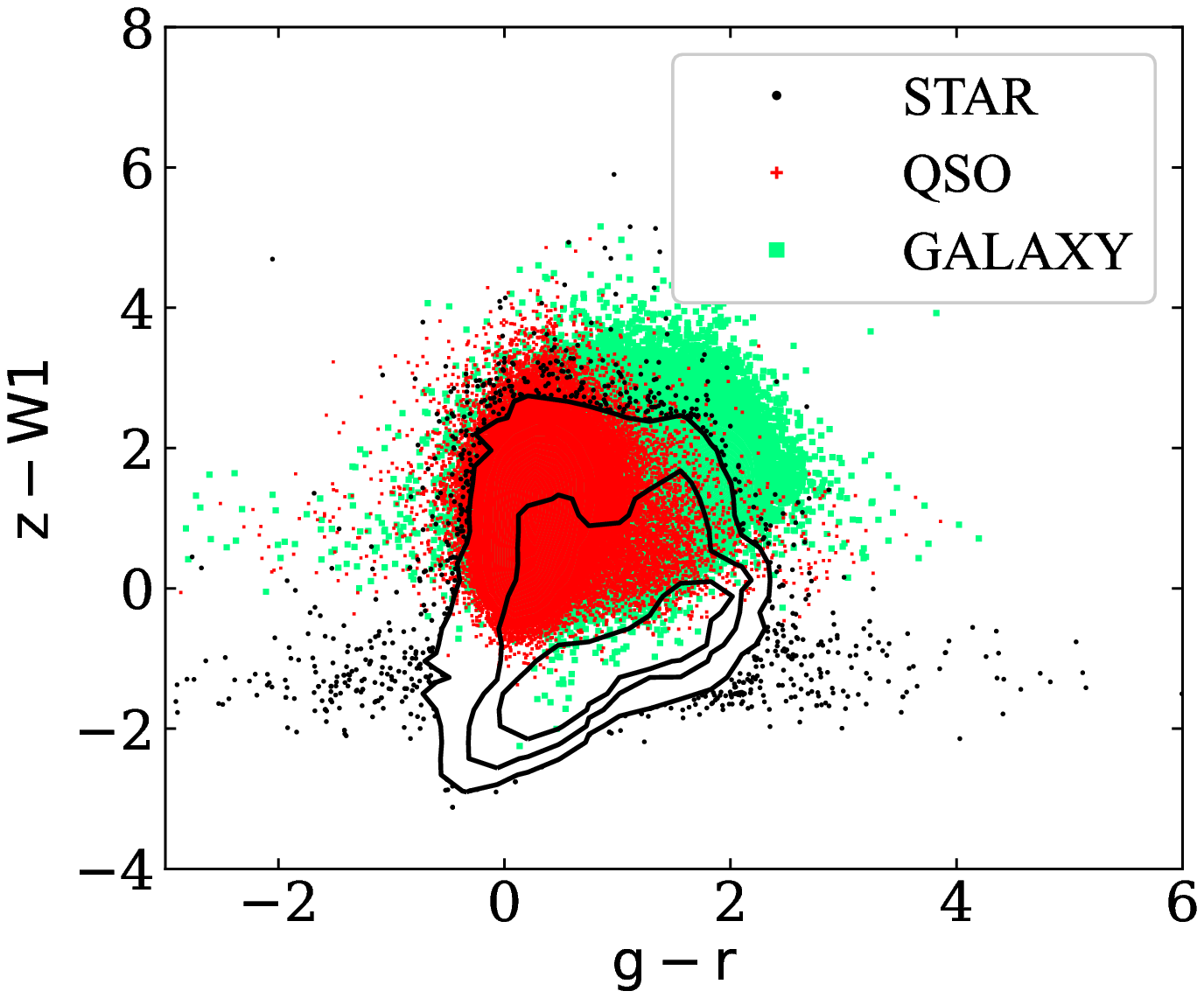}
	\includegraphics[height=5.5cm,width=5.5cm]{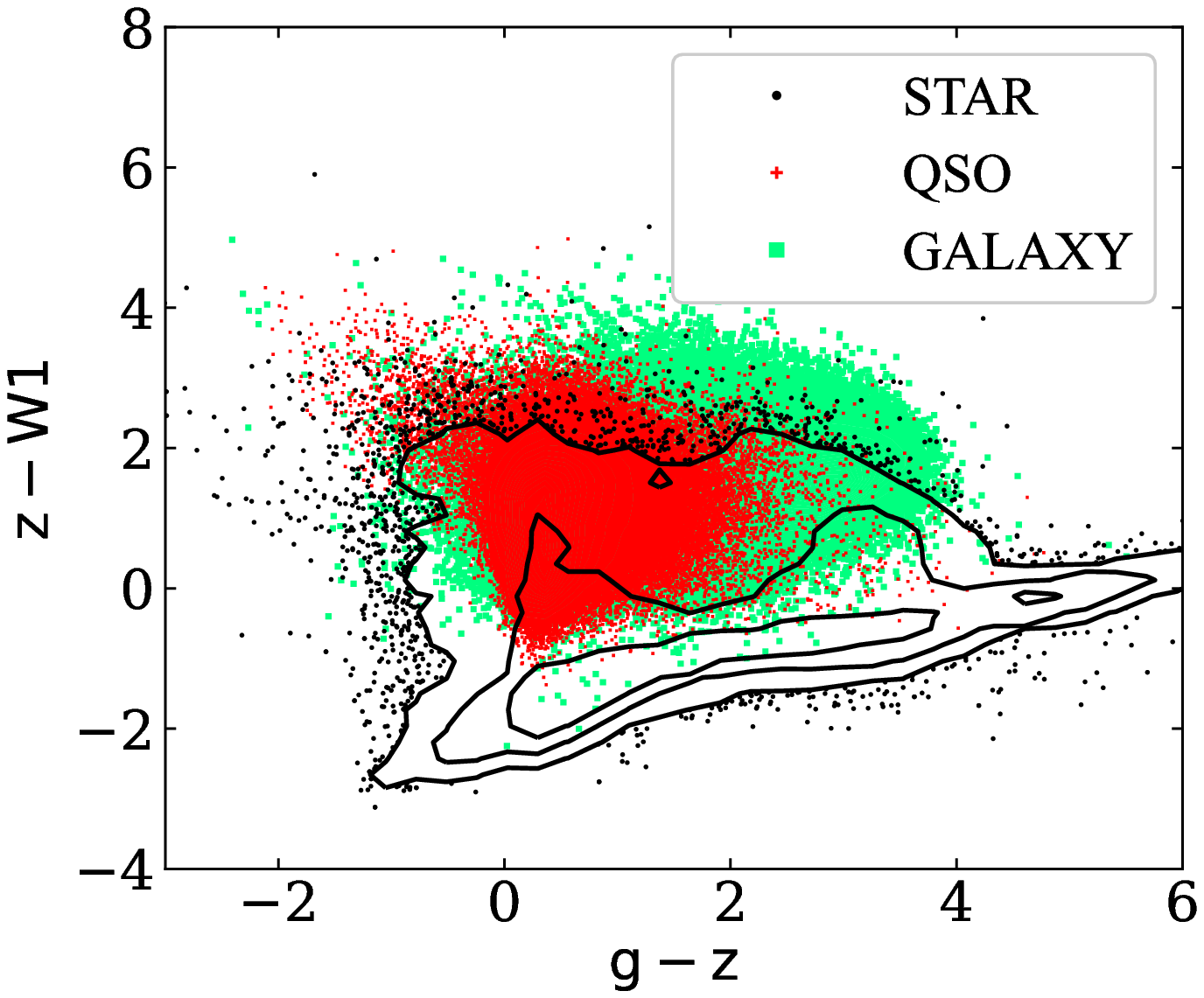}
	\includegraphics[height=5.5cm,width=5.5cm]{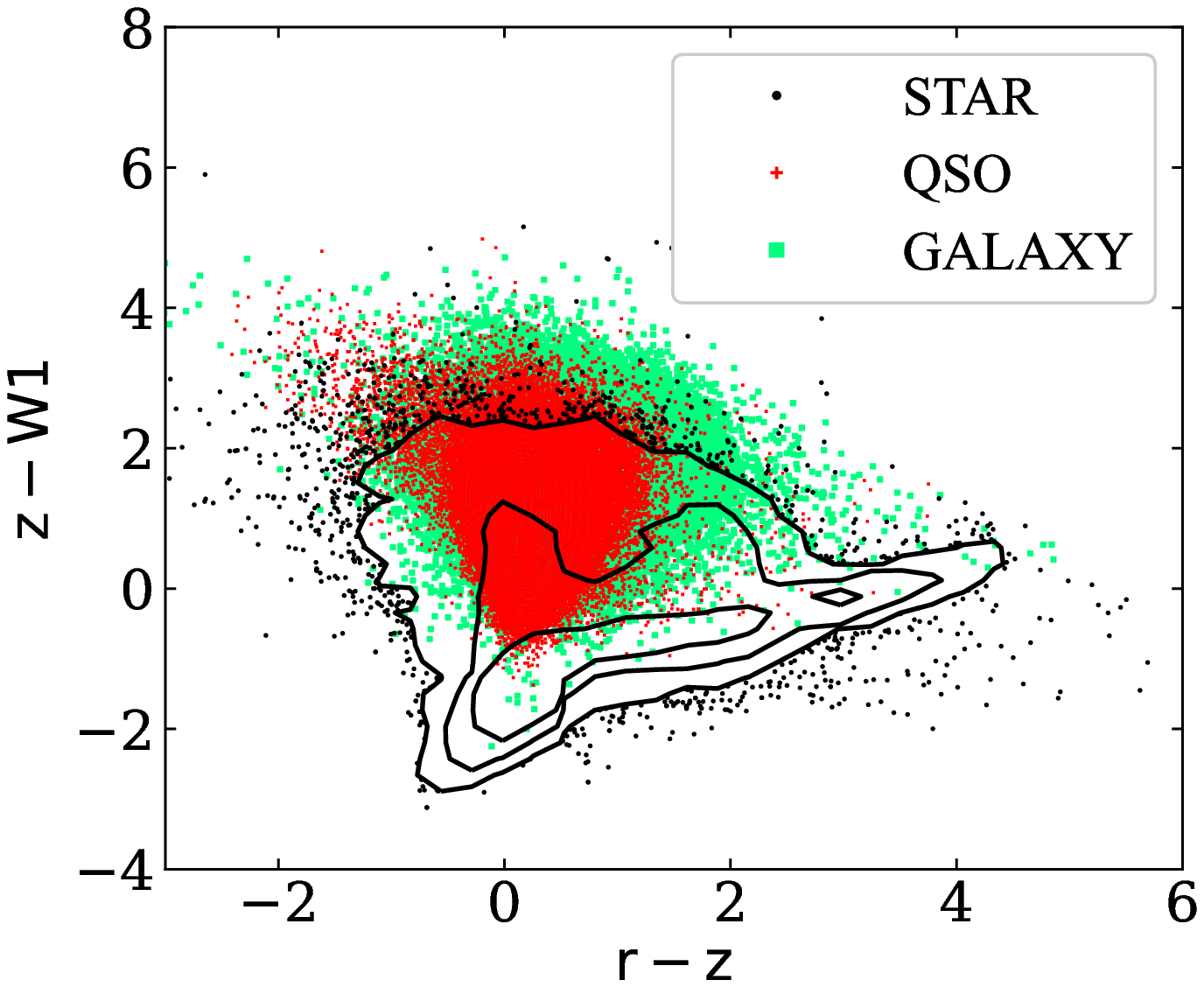}
	\includegraphics[height=5.5cm,width=5.5cm]{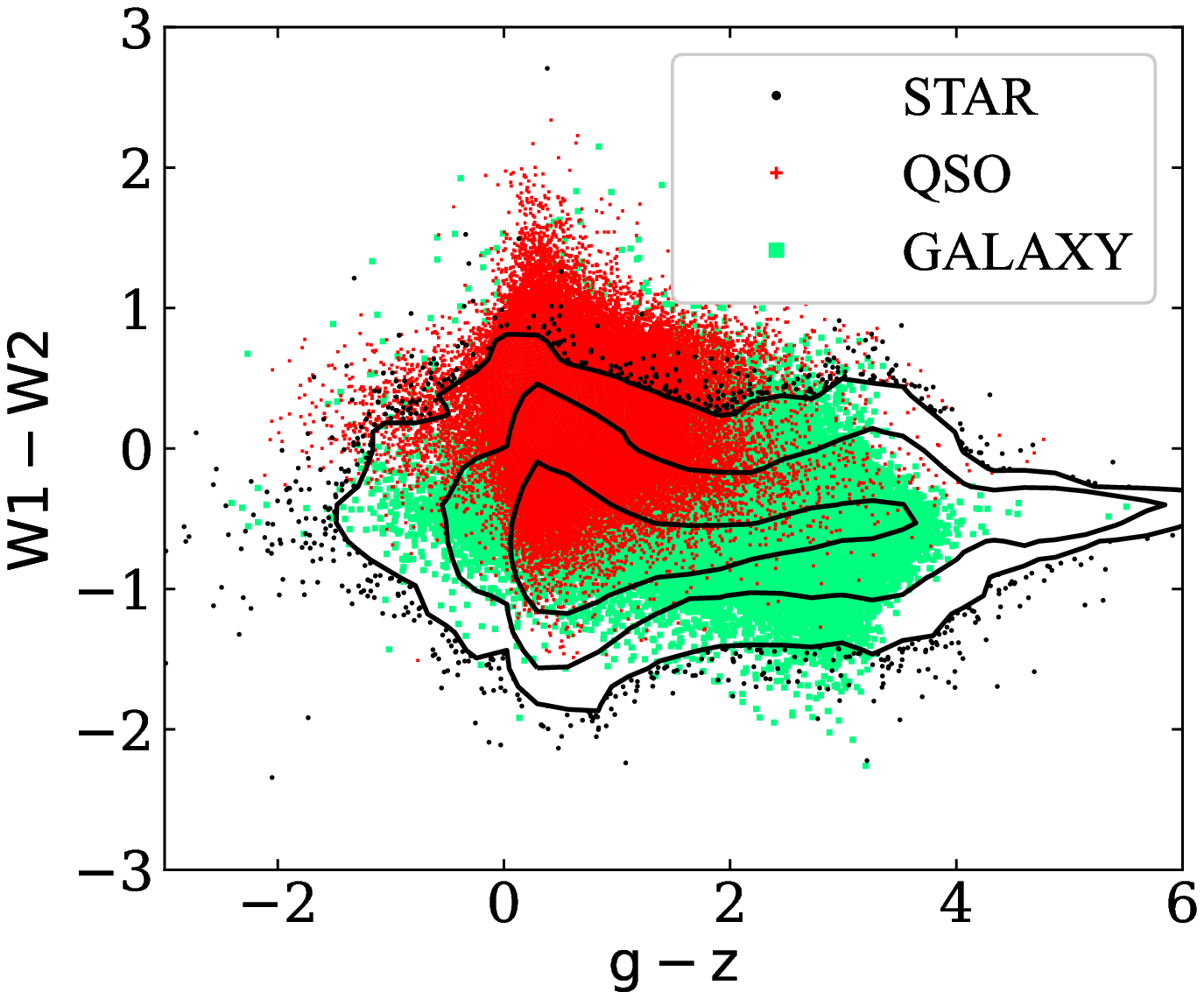}
	\includegraphics[height=5.5cm,width=5.5cm]{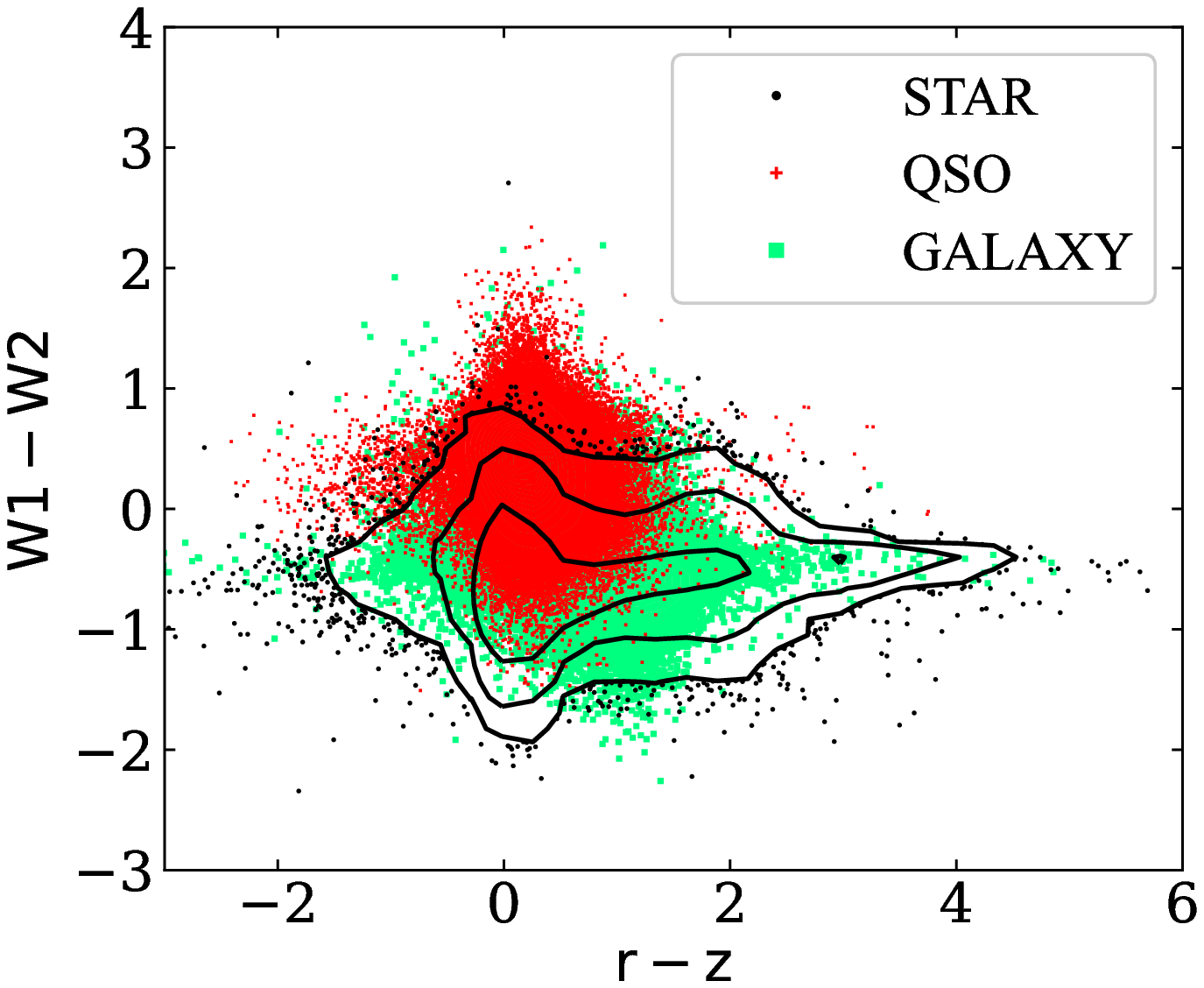}
    \includegraphics[height=5.5cm,width=5.5cm]{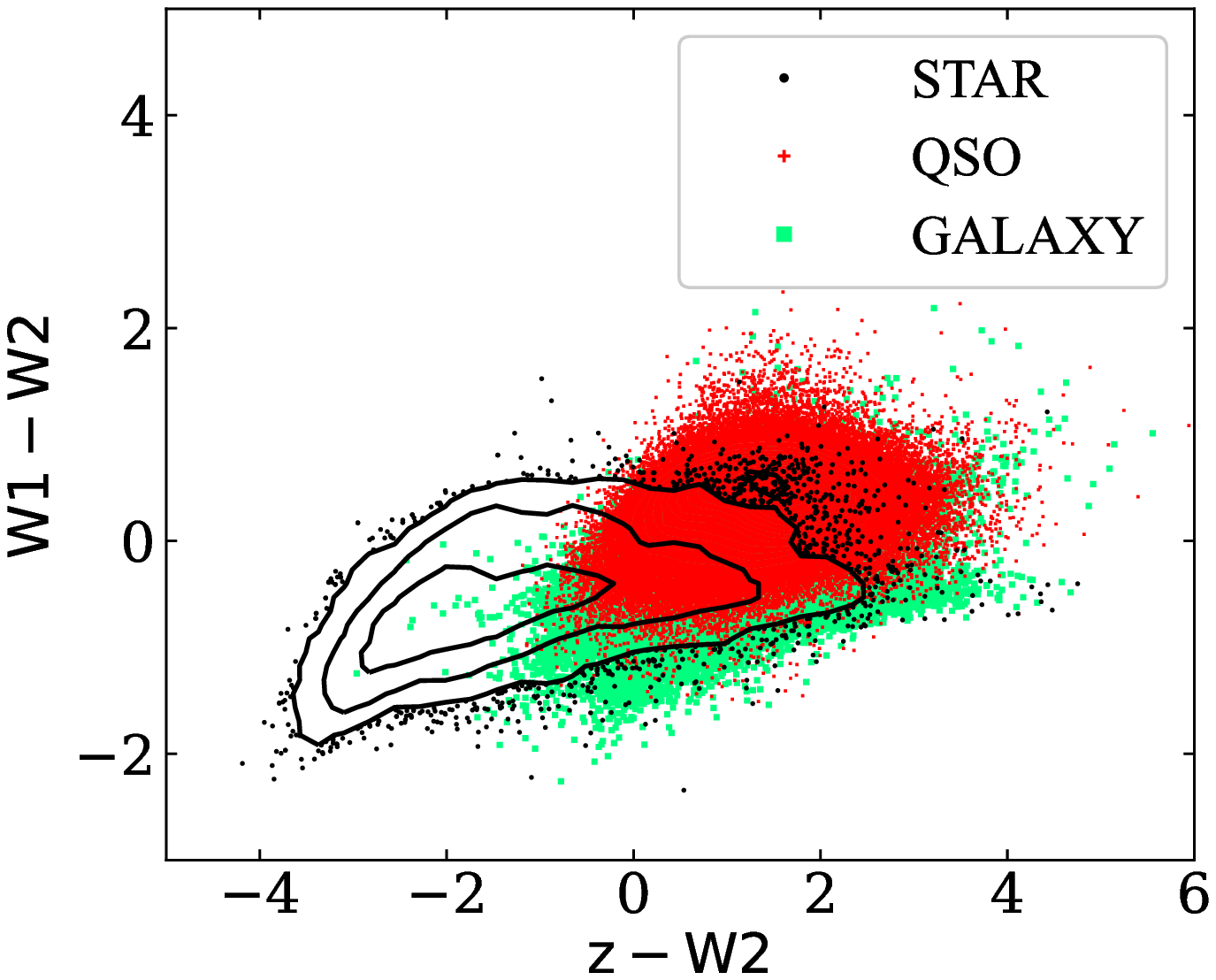}
	\caption{The distribution of stars, galaxies and quasars in 2-d spaces, green filled squares represent galaxies, red pluses represent quasars and black filled circles represent stars. The black outline is the contour of star distribution.}
	\label{fig4}
\end{figure*}

\section{The machine learning model} \label{sec:method}

\subsection{Introduction to XGBoost}

XGBoost \citep{Chen2016} is a boosting algorithm developed on the basis of Gradient Boosting Decision Tree (GBDT) \citep{Friedman2001}. It can solve classification and regression problems. Both XGBoost and GBDT apply boosting method to build strong classifiers by means of learning multiple weak classifiers. But XGBoost has an obvious difference from GBDT. When calculating the residual, the first derivative of the loss function is only used for GBDT, while both the first derivative and the second derivative of the loss function are applied for XGBoost. The objective function of XGBoost can be written in the form of Taylor expansion.
\begin{equation}
Obj^{(t)} = \sum_{i=1}^N [g_{i}f_{t}(x_{i}) + \frac{1}{2} h_{i}f_{t}^{2}(x_{i})] + \Omega(f_{t})
\end{equation}
where $g_{i}$ is the first derivative of the loss function, $h_{i}$ is the second derivative of the loss function.
The term $\Omega$ defines the complexity of the tree, it can be expressed as
\begin{equation}
\Omega(f_{t}) = \gamma T + \frac{1}{2}\lambda\sum_{k=1}^{T}\omega_{k}^{2}
\end{equation}
Here $T$ represents the number of leaves, $\omega_{k}$ represents the score given by the $j$th leaf. The learning of the set of functions used in the model is done by minimizing the objective function. Equation~(2) helps to smooth the final learnt weights to avoid over-fitting.

For a given data set with $n$ examples and $m$ features $\mathcal{D}=\{(x_{i},y_{i})\} (\left|\mathcal{D}\right|={n}, x_{i}\in \mathbb{R}^{m}, y_{i}\in \mathbb{R})$.
Assumed that the model has $k$ trees totally. The prediction on $x_{i}$ is given by
\begin{equation}
\widehat{y_{i}} = \sum_{k=1}^K {f_{k}}(x_{i}),\quad {f_{k}}\in \mathbb{F}
\end{equation}
where $f_{k}$ represents one regression tree, and $f_{k}(x_{i})$ is the score that it gives to $x_{i}$. $\mathbb{F}$ is the space of regression trees.

The importance of the XGBoost algorithm has been widely recognized in a number of machine learning and data mining challenges. Recently XGBoost has wide applications in astronomy, such as classification of unknown source in the Fermi-LAT catalogue \citep{Mirabal2016}, separation of pulsar signals from noise \citep{Bethapudi2018}, and quasar candidate selection \citep{Jinx2019}. In this work, we use XGBoost as a supervised learning algorithm to classify galaxies, stars, and quasars from BASS DR3 sources, and XGboost python package was provided by scikit-learn \citep{scikit-learn}.

\subsection{Classification metrics}

There are many criteria to score the performance of a classifier. Here, we only apply three standard metrics (Accuracy, Precision and Recall) to determine which classifier is better. Accuracy (short for Accu.) is defined as the fraction of the total number of correct predictions among the total number of predictions, Precision (short for Prec.) is the ratio of true positive predictions to all predicted positive examples, Recall (short for Rec.) is the ratio of true positive predictions to all true positive examples. The confusion matrix is a situation analysis table that summarizes the prediction results of the classification model in machine learning, and describes the data in a matrix form according to the two criteria of the real category and the classification judgement made by the classification model. The confusion matrix is suitable for binary and multiclass classification. Through the confusion matrix, Accuracy, Precision and Recall for each class can be calculated. For our case, how to calculate Accuracy, Precision, Recall for triple classification in a confusion matrix is shown in Table~3.

\begin{table*}
\begin{center}
\caption[]{Confusion Matrix \label{tab:confusion}}
 \begin{tabular}{cccccc}
 \hline
Known$\downarrow$Classified$\to$   &GALAXY&QSO&STAR&Precision&Recall\\
\hline
\rule{0pt}{13pt}GALAXY  &$TG$&$FGQ$&$FGS$ &$\displaystyle\frac{TG}{TG+FQG+FSG}$ &$\displaystyle\frac{TG}{TG+FGQ+FGS}$\\
\rule{0pt}{13pt}QSO     &$FQG$&$TQ$&$FQS$ &$\displaystyle\frac{TQ}{TQ+FGQ+FSQ}$ &$\displaystyle\frac{TQ}{TQ+FQG+FQS}$ \\
\rule{0pt}{13pt}STAR    &$FSG$&$FSQ$&$TS$ &$\displaystyle\frac{TS}{TS+FGS+FQS}$ &$\displaystyle\frac{TS}{TS+FSG+FSQ}$ \\
\hline
\rule{0pt}{13pt}Total Accuracy&\multicolumn{5}{c}{$\displaystyle\frac{TG+TQ+TS}{TG+FGQ+FGS+TQ+FQG+FQS+TS+FSG+FSQ}$}\\
\hline
\multicolumn{6}{l}{$^a$ TG  represents the number of actual galaxies that are correctly classified.}\\
\multicolumn{6}{l}{$^b$ FGQ represents the number of galaxies that are incorrectly classified as quasars.}\\
\multicolumn{6}{l}{$^c$ FGS represents the number of galaxies that are incorrectly classified as stars.}\\
\multicolumn{6}{l}{$^d$ TQ  represents the number of actual quasars that are correctly classified.}\\
\multicolumn{6}{l}{$^e$ FQG represents the number of quasars that are incorrectly classified as galaxies.}\\
\multicolumn{6}{l}{$^f$ FQS represents the number of quasars that are incorrectly classified as stars.}\\
\multicolumn{6}{l}{$^g$ TS  represents the number of actual stars that are correctly classified.}\\
\multicolumn{6}{l}{$^h$ TSG represents the number of stars that are incorrectly classified as galaxies.}\\
\multicolumn{6}{l}{$^i$ TSQ represents the number of stars that are incorrectly classified as quasars.}\\
\end{tabular}
\end{center}
\end{table*}

\section{The Classifier Model} \label{sec:performance}

We randomly divide the known sample into 10 equal subsets keeping the same proportion of the three classes (stars, galaxies and quasars). When training a classifier, we adopt 10-fold cross-validation, which means that 9 subsets are used as training set and the left subset is taken as test set in turn, then the average Accuracy, Precision and Recall are calculated for ten experiments. This work uses XGBoost's binary classifier (classification of galaxies from stars and quasars) and multiclass classifier (classification of galaxies, stars and quasars) respectively. There are two steps to create classifiers. The first step is to get feature importance by XGBoost, then we try different input patterns of different samples to train XGBoost models and get the best model parameters of XGBoost by grid search. The three main hyper parameters used for XGBoost model optimization include the number of trees in the forest ($n\_estimators$), Maximum depth of a tree ($max\_depth$), Step size shrinkage used in update to prevents overfitting ($learning\_rate$). All other hyper parameters are set with their default values. The second step is to evaluate the performance of different classifiers by ten-fold validation with different input patterns. Finally the best classifiers for different samples are created and then can be applied for prediction of unknown sources. All computing was run in the cloud computing environment of National Astronomical Data Centre (NADC) \citep{ Li2017}.

\subsection{Feature importance}
The input pattern is a importance factor to affect the performance of XGBoost. XGBoost may rank relative importance of each feature when used for feature importance analysis according to the gain, which implies the relative contribution of a feature to the XGBoost model. Compared to another feature, the higher gain value of a feature, the more important it is for providing a prediction. Firstly, we evaluate the importance of all possible attributes (${\Delta}g$, ${\Delta}r$, ${\Delta}z$, $g$, $r$, $z$, $g-r$, $r-z$, $g-z$) for optical sample (Sample~I) by XGBoost when separating point sources and extended sources (Figure~3 panel~(A)) and when discriminating quasars and stars (Figure~3 panel~(B)). Secondly, we also assess the importance of all possible attributes (${\Delta}g$, ${\Delta}r$, ${\Delta}z$, $g$, $r$, $z$, $W1$, $W2$, $g-r$, $r-z$, $g-z$, $z-W1$, $g-W1$, $r-W1$, $z-W2$, $g-W2$, $r-W2$, $W1-W2$) for optical and infrared sample (Sample~II) by XGBoost when separating point sources and extended sources (Figure~3 panel~(C)) and when discriminating quasars and stars (Figure~3 panel~(D)). Figure~3 shows that the feature importance is related to samples and classification task. When classifying point sources and extended sources, the rank of feature importance for Sample~I is ${\Delta}z$, ${\Delta}r$, ${\Delta}g$, $g-r$, $g$, $g-z$, $r$, $z$, $r-z$; the rank for Sample~II is ${\Delta}g$, ${\Delta}z$, $g-W1$, $W1-W2$, $z-W1$, ${\Delta}r$, $g-z$, $z-W2$, $g-r$, $r-z$, $W1$, $r$, $g$, $z$, $r-W2$, $W2$, $r-W1$, $g-W2$. While separating quasars and stars, the rank of feature importance for Sample~I is $r$, $g-z$, $z$, $g-r$, ${\Delta}z$, $r-z$, ${\Delta}r$, ${\Delta}g$, $g$; the rank for Sample~II is $z-W2$, $W1-W2$, $g-z$, $g-r$, $z-W1$, $r-z$, ${\Delta}z$, $r$, $r-W2$, $z$, ${\Delta}g$, $g-W1$, ${\Delta}r$, $W1$, $g-W2$, $g$, $r-W1$, $W2$. The feature importance is helpful to select effective features for classification. Figure~3 also shows that ${\Delta}g$, ${\Delta}z$, ${\Delta}r$ are very important for distinguishing extended sources from point sources, which is consistent with Figure~1. As shown in Figure~3, $r$, $g-z$, ${\Delta}z$ and ${\Delta}g$ of all features are more important for classification of stars and quasars with optical information; $z-W2$, $W1-W2$, $g-z$ and $g-r$ of all features contribute more to classification of stars and quasars with combined optical and infrared information.

\begin{figure*}
	\centering
	\includegraphics[height=6.5cm,width=8.5cm]{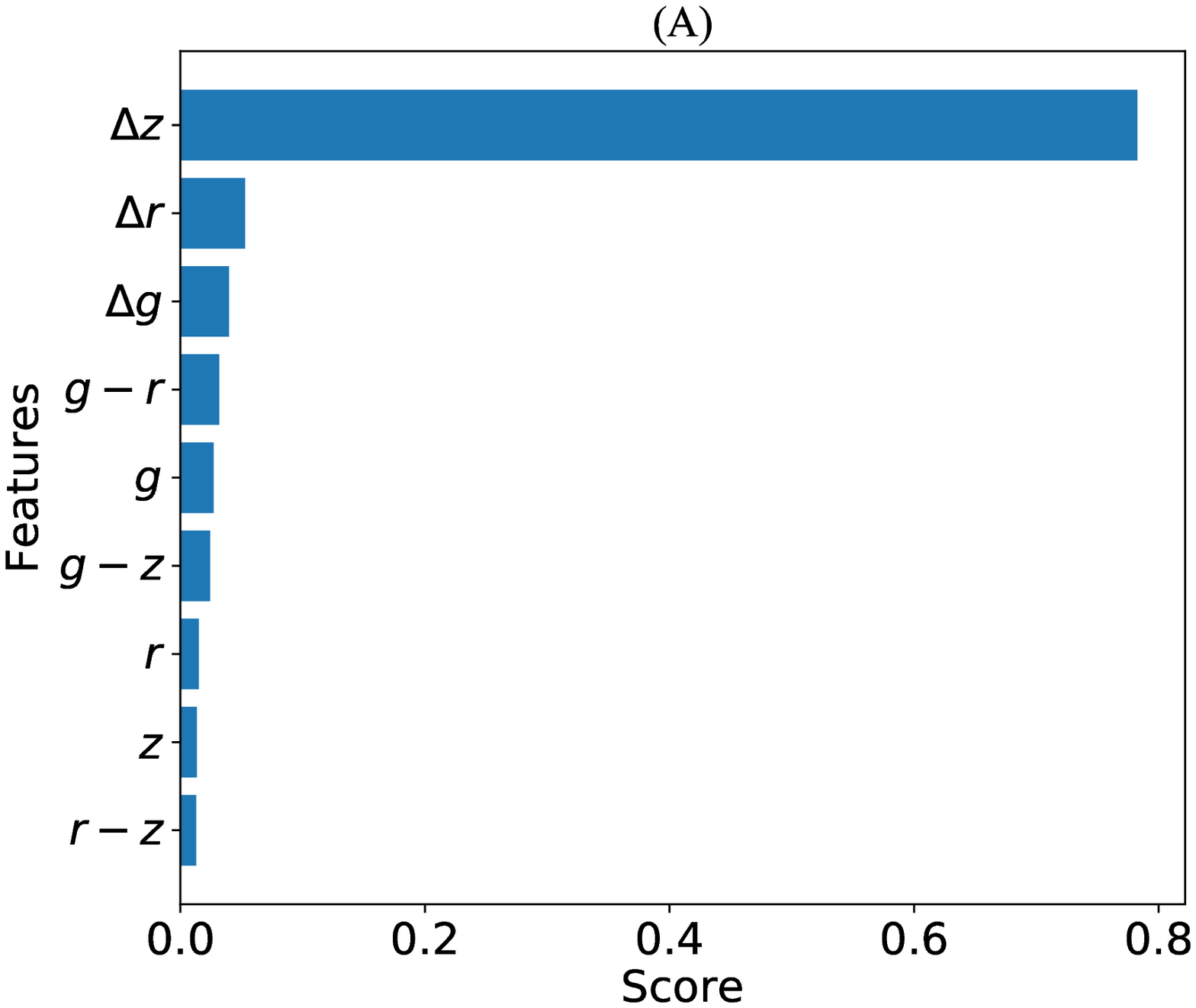}
	\includegraphics[height=6.5cm,width=8.5cm]{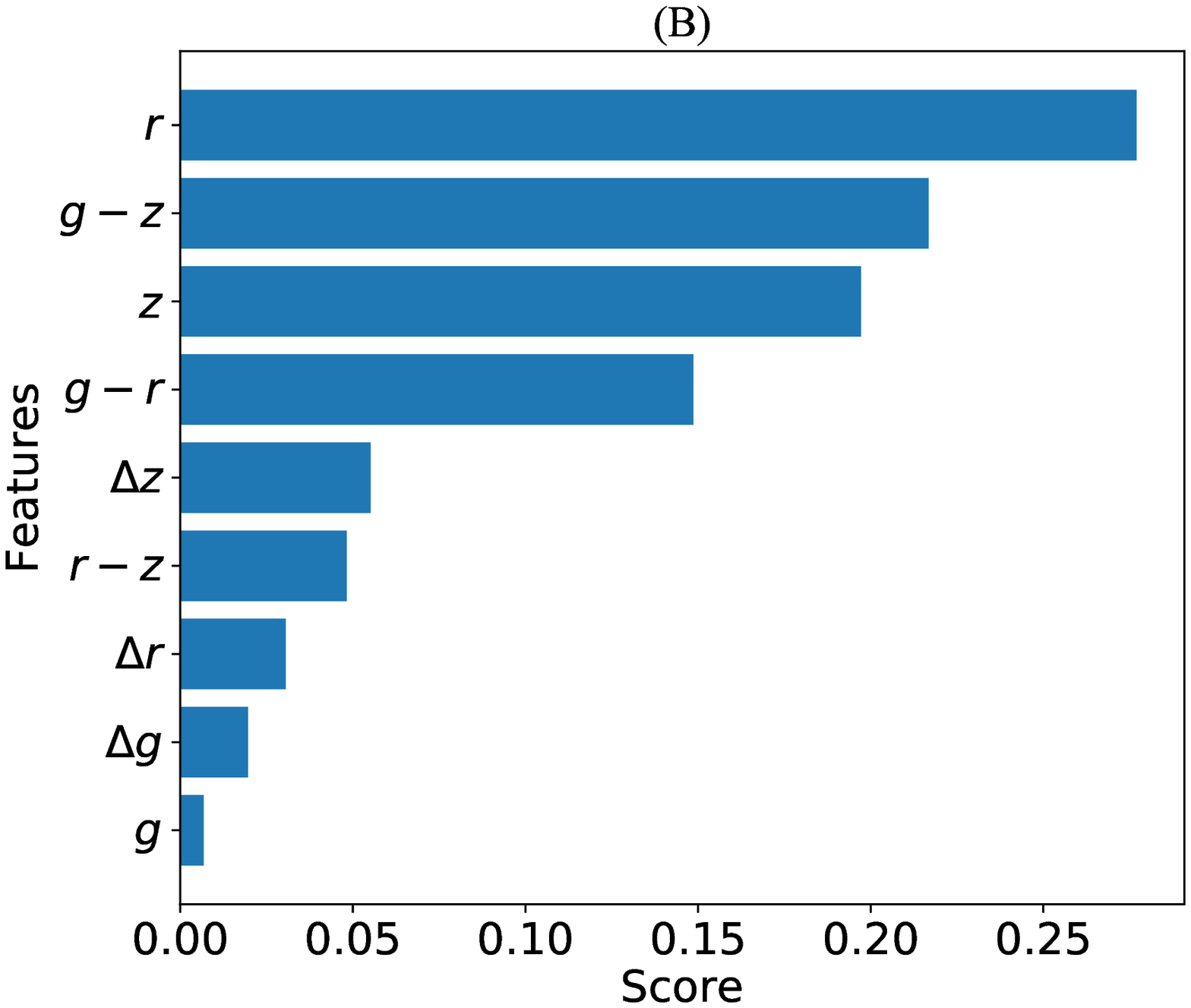}
	\includegraphics[height=6.5cm,width=8.5cm]{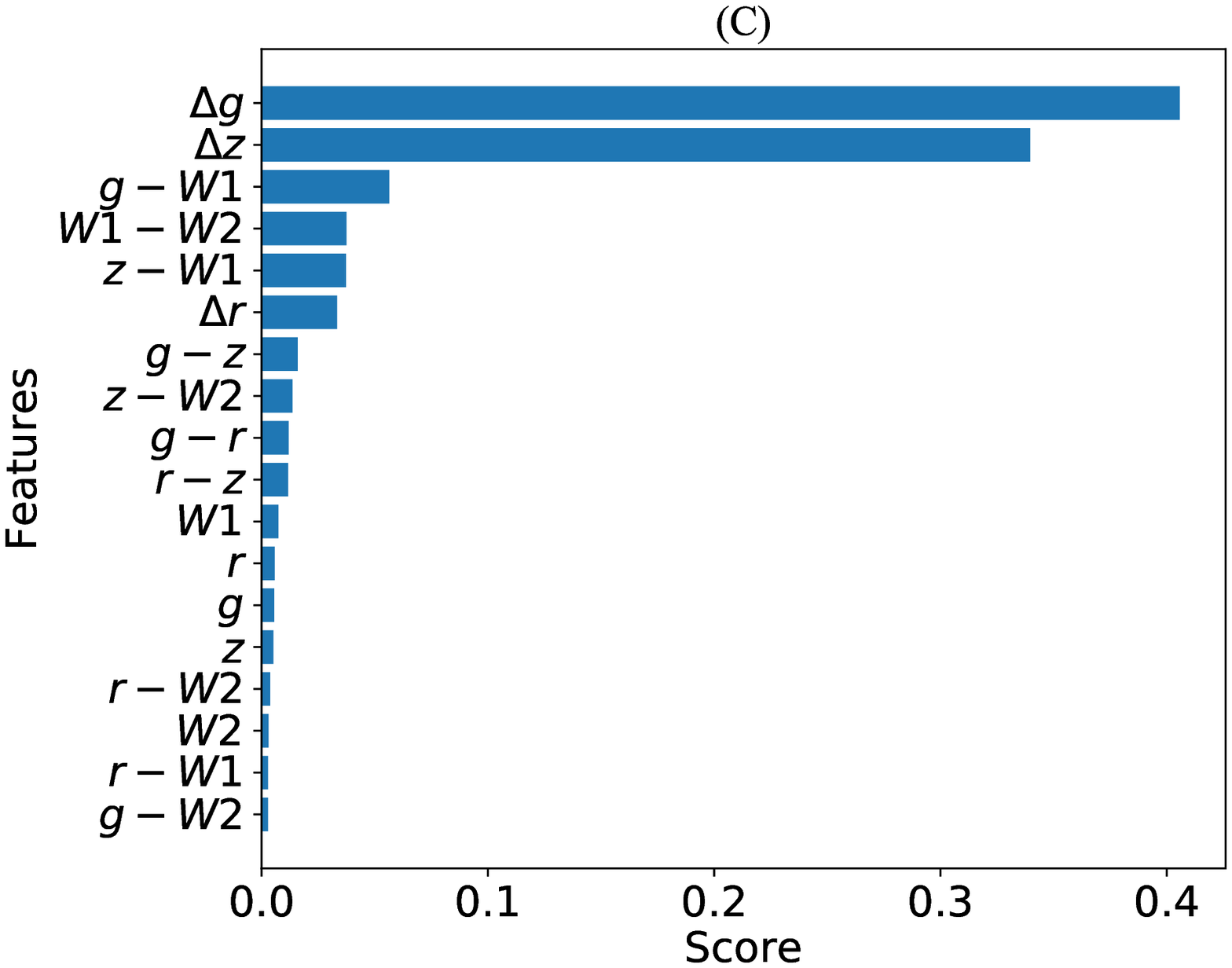}
	\includegraphics[height=6.5cm,width=8.5cm]{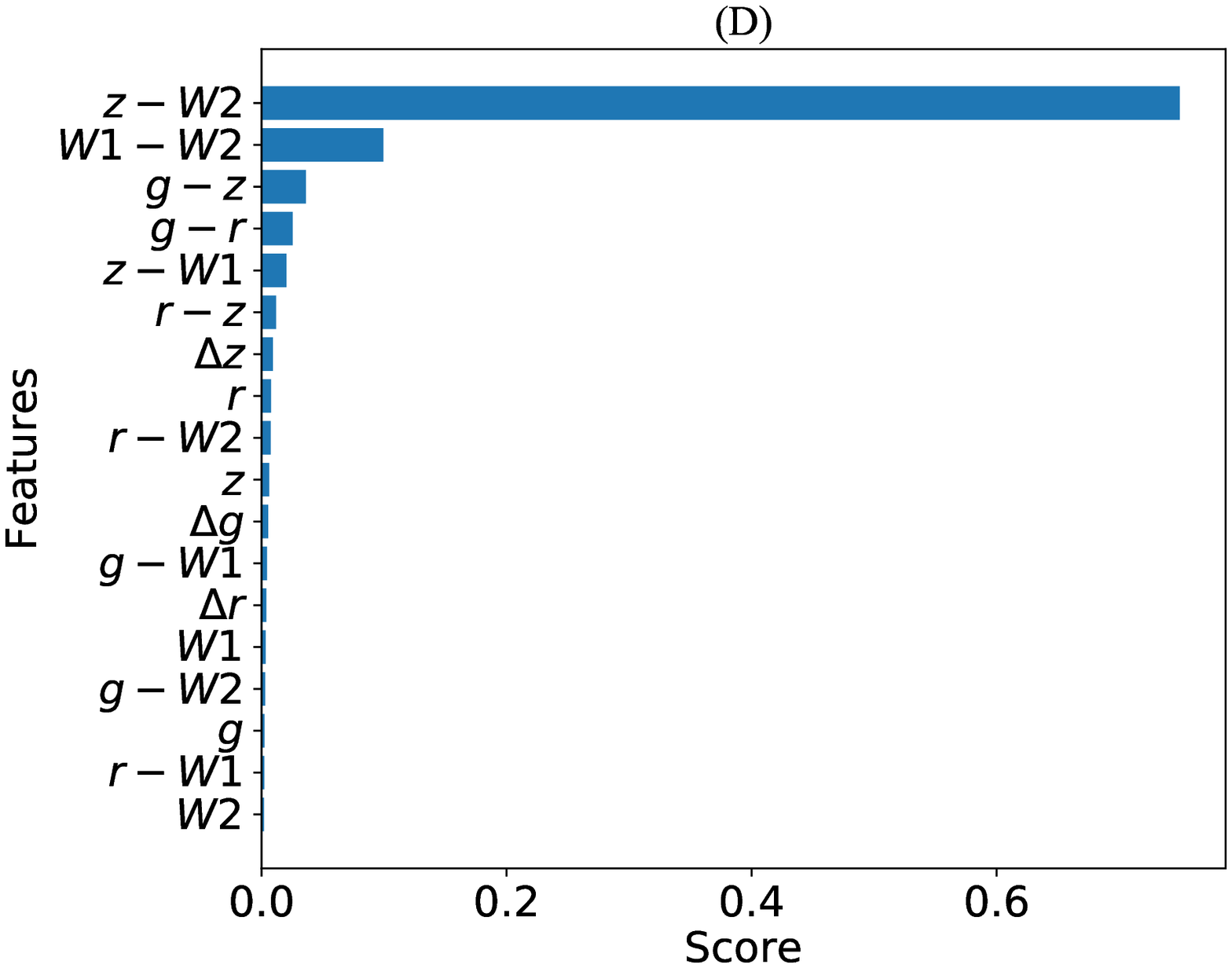}
	\caption{The feature importance with XGBoost. Panel~(A) shows the feature importance only with the optical information when classifying extended sources and point sources; panel~(B) shows the feature importance only with the optical information when classifying stars and quasars; panel~(C) shows the feature importance with combined optical and infrared information when classifying extended sources and point sources; panel~(D) shows the feature importance with combined optical and infrared information when classifying stars and quasars. }	
	\label{fig3}
\end{figure*}

\subsection{Binary classifier construction of XGBoost}
We firstly distinguish extended sources (galaxies) from point sources (stars and quasars) by a binary XGBoost classifier. Then the point sources are further classified by another binary XGBoost classifier, and finally the three class labels are assigned to the predicted sample, i.e. we apply two-layer classifiers to realize the classification of galaxies, stars and quasars. After training by grid-search method with all features, we get the best model parameters: $max\-depth=11$, $n\_estimators=100$ and $learning\_rate=0.5$. For the classification of point and extended sources, the classification performance obtained by the different input patterns is shown in Table~4. As indicated in Table~4, the best input pattern is 8-features (${\Delta}g$, ${\Delta}r$, ${\Delta}z$, $g-r$, $g-z$, $r$, $r-z$, $g$) only with optical information for XGBoost. Accuracy is 97.28 per cent, Precision and Recall are more than 96.60 per cent for point sources and extended sources. The optimal input pattern is ${\Delta}g$, ${\Delta}z$, $g-W1$, $W1-W2$, $z-W1$, ${\Delta}r$, $g-z$, $z-W2$, $g-r$,$r-z$,$W1$, $r$, $g$, $z$, $r-W2$ with both optical and infrared information for XGBoost. Accuracy amounts to 98.67 per cent, both Precision and Recall are more than 98.30 per cent for point sources while Precision and Recall are larger than 98.80 per~cent for extended sources. Therefore it is reliable to separate extended sources from point sources by XGBoost with optical information or combined optical and infrared information. Moreover it is evident that adding infrared information is helpful to improve classification performance.

\begin{table*}
\begin{center}
\caption[]{The performance of binary classifier for point and extended sources \label{tab:comp2}}
 \begin{tabular}{rccccccc}
 \hline
  & && \multicolumn{2}{c}{Point Sources}  & \multicolumn{2}{c}{Extended Sources}  & \\
 \hline
Input pattern  &Sample &  Accu.(\%)  &Prec.(\%)& Rec.(\%)&Prec.(\%)&Rec.(\%)&Time(s)\\
\hline
(${\Delta}g$, ${\Delta}r$, ${\Delta}z$)                     &Sample~I   &95.72&94.98&95.75&96.36&95.69&21\\
(${\Delta}g$, ${\Delta}r$, ${\Delta}z$, $g-r$, $r-z$)       &Sample~I     &96.73&96.07&96.84&97.29&96.63&34 \\
(${\Delta}g$, ${\Delta}r$, ${\Delta}z$, $g-r$, $r-z$, $g-z$)&Sample~I     &96.78&96.08&96.96&97.39&96.63&47 \\
(${\Delta}g$, ${\Delta}r$, ${\Delta}z$, $g-r$, $r-z$, $r$)  &Sample~I     &97.22&96.57&97.40&97.77&97.06&54 \\
(\texttt{8-features})   &Sample~I    &\textbf{97.28} &\textbf{96.63} &\textbf{97.47} &\textbf{97.83} &\textbf{97.11}& 73 \\
(\texttt{11-features})  &Sample~II   &98.52&98.26&98.11&98.70&98.80&90\\
(\texttt{13-features})  &Sample~II   &98.64&98.35&98.32&98.84&98.86&105 \\
(\texttt{15-features})  &Sample~II   &\textbf{98.67}&\textbf{98.39}&\textbf{98.35}&\textbf{98.86}&\textbf{98.89}&111 \\
(\texttt{18-features}) &Sample~II   &98.65 &98.37 &98.32 &98.84 &98.87 &152 \\
 \hline
 \multicolumn{8}{l}{$^a$ \texttt{8-features} represents ${\Delta}g$, ${\Delta}r$, ${\Delta}z$, $g-r$, $g-z$, $r$, $r-z$, $g$. }\\
 \multicolumn{8}{l}{$^b$ \texttt{11-features} represents ${\Delta}g$, ${\Delta}r$, ${\Delta}z$, $g-r$, $r-z$, $g-z$, $g$, $r$, $z$, $W1$, $W2$. }\\
 \multicolumn{8}{l}{$^c$ \texttt{13-features} represents ${\Delta}g$, ${\Delta}z$, $g-W1$, $W1-W2$, $z-W1$, ${\Delta}r$, $g-z$, $z-W2$, $g-r$,$r-z$,$W1$, $r$, $g$. }\\
 \multicolumn{8}{l}{$^d$ \texttt{15-features} represents ${\Delta}g$, ${\Delta}z$, $g-W1$, $W1-W2$, $z-W1$, ${\Delta}r$, $g-z$, $z-W2$, $g-r$,$r-z$,$W1$, $r$, $g$, $z$, $r-W2$. }\\
\end{tabular}
\end{center}
\end{table*}

Then we train the XGBoost classifier on the point sources (stars and quasars), and the classification performance is shown in Table~5. As described in Table~5, the best input pattern is ${\Delta}g$, ${\Delta}r$, ${\Delta}z$, $g-r$, $r-z$, $g-z$, $r$, $g$, $z$ only depending on optical information for XGBoost. Accuracy is 93.22 per cent, Precision and Recall are respectively 93.33 per cent and 93.71 per cent for stars while Precision and Recall are respectively 93.11 per cent and 92.69 per cent for quasars. The optimal input pattern is $z-W2$, $W1-W2$, $g-z$, $g-r$, $z-W1$, $r-z$, ${\Delta}z$, $r$, $r-W2$, $z$, ${\Delta}g$, $g-W1$, ${\Delta}r$, $W1$, $g-W2$ with both optical and infrared information for XGBoost. Accuracy amounts to 99.15 per cent, Precision and Recall are respectively 99.00 per cent and 99.46 per cent for stars while Precision and Recall are respectively 99.33 per cent and 98.77 per cent for quasars. In terms of Accuracy, Precision and Recall, XGBoost classifiers are effective to separate quasars from stars with optical information or combined optical and infrared information. As expected, adding infrared information contributes to classification performance.

\begin{table*}
\begin{center}
\caption[]{The performance of binary classifier for stars and quasars by XGBoost. \label{tab:comp3}}
 \begin{tabular}{rccccccc}
 \hline
  & &&\multicolumn{2}{c}{STAR} &\multicolumn{2}{c}{QSO} & \\
 \hline\noalign{\smallskip}
Input pattern  &Sample &  Accu.(\%)  &Prec.(\%)& Rec.(\%)&Prec.(\%)&Rec.(\%)&Time(s)\\
\hline
(${\Delta}g$, ${\Delta}r$, ${\Delta}z$, $g-r$)  &Sample~I   &84.68&83.13&88.62&86.62&80.38&12\\
($r$, $g-z$,$z$,$g-r$)  &Sample~I   &91.31&91.40&91.99&91.20&90.56&17\\
($r$, $g-r$,$r-z$)  &Sample~I   &91.75&91.71&92.55&91.79&90.87&12\\
(${\Delta}g$, ${\Delta}r$, ${\Delta}z$, $g-r$, $r-z$, $r$)  &Sample~I   &93.20&93.25&\textbf{93.75}&\textbf{93.14}&92.60&25 \\
(\texttt{9-features})     &Sample~I  &\textbf{93.22}&\textbf{93.33}&93.71&93.11&\textbf{92.69}&37 \\
($g-r$, $r-z$, $r$, $g$, $z$, $w1$, $w2$) &Sample~II     &98.88&98.61&99.38&99.23&98.27&24\\
(\texttt{10-features})     &Sample~II  &99.11&98.95&99.45&99.32&98.70&29 \\
(\texttt{12-features})     &Sample~II &99.13&98.92&99.51&99.40&98.67&17 \\
(\texttt{15-features})     &Sample~II  &\textbf{99.15} &\textbf{99.00} &99.46 &99.33 &\textbf{98.77} &25 \\
(\texttt{17-features}) &Sample~II &99.12&98.94&\textbf{99.47}&\textbf{99.35}&98.69&36 \\
\hline
\multicolumn{8}{l}{$^a$ \texttt{9-features} represents ${\Delta}g$, ${\Delta}r$, ${\Delta}z$, $g-r$, $r-z$, $g-z$, $r$, $g$, $z$.}\\
\multicolumn{8}{l}{$^b$ \texttt{10-features} represents $z-W2$,$z-W1$, $g-r$, $W1-W2$, ${\Delta}z$,${\Delta}g$, ${\Delta}r$, $r-z$, $r-W2$, $r$.}\\
\multicolumn{8}{l}{$^c$ \texttt{12-features} represents $z-W2$,$z-W1$, $g-z$, $W1-W2$, $g-r$, ${\Delta}z$, $r-z$, $r-W2$, $r$, $z$, ${\Delta}g$, $g-W1$.}\\
\multicolumn{8}{l}{$^d$ \texttt{15-features} represents $z-W2$, $W1-W2$, $g-z$, $g-r$, $z-W1$, $r-z$, ${\Delta}z$, $r$, $r-W2$, $z$, ${\Delta}g$, $g-W1$, ${\Delta}r$, $W1$, $g-W2$.}\\
\multicolumn{8}{l}{$^e$ \texttt{17-features} represents $z-W2$,$z-W1$, $g-z$, $W1-W2$, $g-r$, ${\Delta}z$, $r-z$, $r-W2$, $r$, $z$, ${\Delta}g$, $g-W1$,  ${\Delta}r$, $g$, $g-W2$, $W1$, $W2$.}\\
\end{tabular}
\end{center}
\end{table*}

According to the above experimental results, the difference between $PSFMag$ and $KronMag$ in $g$, $r$, and $z$ bands has great impact on the distinction between point and extended sources, while for the classification of stars and quasars, the infrared features $W1$, $W2$ are very important. When using two-layer binary classifiers, the first-layer classification performance will also affect the second-layer classification prediction. Considering running time, XGBoost is a fast algorithm for classification and shows its superiority in dealing with large scale data.

\subsection{Multiclass classifier construction of XGBoost}

Compared with two-layer binary classifiers, it is simple to use multiclass classifier to classify galaxies, stars and quasars. For multiclass classifier, you only need to set $objective$ to $multi:softmax$ and $num\_class$ to 3 in the XGBoost model. The basic training process is exactly the same as the binary classifier model training. We use the grid search method to get the optimal model parameters and 10-fold cross validation to get average Accuracy, Precision and Recall values. The best model parameters for XGBoost are $max\-depth=7$, $n\_estimators=200$ and $learning\_rate=0.5$ According to the confusion matrix, we calculate the classification Accuracy, Precision and Recall for each class. Table~6 shows the performance of multi-class classifiers with four kinds of input patterns for classification of galaxies, stars and quasars. As shown in Table~6, there are two input patterns for the samples only with optical information: input pattern I for ${\Delta}g$, ${\Delta}r$, ${\Delta}z$, $g-r$, $r-z$, $r$ and input pattern II for ${\Delta}g$, ${\Delta}r$, ${\Delta}z$, $g-r$, $g-z$, $r$, $r-z$, $z$; also two input patterns for the samples with both optical and infrared information, input pattern III for $z-W2$, ${\Delta}z$, $W1-W2$, ${\Delta}r$, $g-r$, $z-W1$, $W1$, $W2$ and input pattern IV for $z-W2$, ${\Delta}z$, $W1-W2$, ${\Delta}r$, $g-r$, $z-W1$, ${\Delta}g$, $g-z$, $r-W2$, $r-z$, $r$. Comparing the performance of input pattern II with that of input pattern I, Accuracy is better (94.49 per cent vs. 94.47 per cent), Precision and Recall are better for stars and galaxies, Recall is better and Precision is a little worse for quasars. Considering all these, the performance of input pattern II is a little superior to that of input pattern I. Comparing the performance of input pattern IV with that of input pattern III, Accuracy is better (98.43 per cent vs. 97.90 per cent), Precision and Recall are all better for stars, galaxies and quasars. Given these, the performance of input pattern IV outperforms that of input pattern III.

\begin{table*}
\begin{center}
\caption[]{ The performance of multi-class classifier for galaxies, stars and quasars. }
\begin{tabular}{rccccccccc}
\hline
          &      &       &\multicolumn{2}{c}{GALAXY}&\multicolumn{2}{c}{STAR}&\multicolumn{2}{c}{QSO}&\\
\hline
 Input pattern&Sample&Accu.(\%)&Prec.(\%)&Rec.(\%)&Prec.(\%)&Rec.(\%)&Prec.(\%)&Rec.(\%)&Time(s)\\
\hline
Input pattern I   &Sample~I   &94.47  &97.74  &97.32  &92.96  &90.59  &\textbf{88.68} &91.63 &326 \\
Input pattern II  &Sample~I   &\textbf{94.49}  &\textbf{97.76}  &\textbf{97.34}  &\textbf{93.00}  &\textbf{90.60}  &88.67 &\textbf{91.65} &260 \\
Input pattern III &Sample~II  &97.90  &98.36  &98.76  &98.32  &95.50  &96.37 &97.62 &160 \\
Input pattern IV  &Sample~II  &\textbf{98.43}  &\textbf{98.85}  &\textbf{98.97}  &\textbf{98.74}  &\textbf{97.28}  &\textbf{97.07} &\textbf{97.95} &330 \\
\hline
\multicolumn{10}{l}{ $^a$ Input pattern I represents ${\Delta}g$, ${\Delta}r$, ${\Delta}z$, $g-r$, $r-z$, $r$.}\\
\multicolumn{10}{l}{ $^b$ Input pattern II represents ${\Delta}g$, ${\Delta}r$, ${\Delta}z$, $g-r$, $g-z$, $r$, $r-z$, $z$.}\\
\multicolumn{10}{l}{ $^c$ Input pattern III represents $z-W2$, ${\Delta}z$, $W1-W2$, ${\Delta}r$, $g-r$, $z-W1$, $W1$, $W2$.}\\
\multicolumn{10}{l}{ $^d$ Input pattern IV represents $z-W2$, ${\Delta}z$, $W1-W2$, ${\Delta}r$, $g-r$, $z-W1$, ${\Delta}g$, $g-z$, $r-W2$, $r-z$, $r$.}\\
\end{tabular}
\end{center}
\end{table*}

\subsection{Discussion}
Our goal is to separate galaxies, stars and quasars with photometric data (BASS optical filters: $g$, $r$, $z$; ALLWISE mid-IR: $W1$, $W2$) using XGBoost. In their optical images, galaxies are easily discriminated from stars and quasars due to their extended morphology. As described in Figure~1, the extended characteristics of galaxies is distinct in ${\Delta}z$, ${\Delta}r$ and ${\Delta}g$ whose absolute values are not zero; stars and quasars are similar and pointed, and the absolute values of ${\Delta}z$, ${\Delta}r$ and ${\Delta}g$ for them are near zero. From Figure~2, it is known that galaxies, stars and quasars are difficult to separate from each other with one or two of all the features although their separation improves a lot with the additional infrared information and their overlapping still exists.

Compared to 1-d histogram and 2-d scatter plot, XGBoost may take all features into account and solve the classification problem. As shown in Tables~4-6, it is concluded that the classification accuracy increases for any classifier when adding the information from infrared band. The best performance is obtained with all information from optical and infrared bands. If one source is predicted both by classifiers with optical information and by those with combined optical and infrared information, the predicted results by classifiers with combined optical and infrared information are more reliable than those by classifiers with optical information, the same predicted results by all these classifiers is the most reliable. As shown in Table~4, Figure~1 and Figure~3, ${\Delta}z$, ${\Delta}r$ and ${\Delta}g$ are of great importance for separation of extended and point sources (Accuracy: more than 95.70 per cent) no matter for Sample~I or Sample~II, the infrared information influence the performance of a classifier to some extent. As indicated in Figure~3 and Table~5, the four important features ($r$, $g-z$, ${\Delta}z$ and ${\Delta}g$) for Sample~I are different from those ($z-W2$, $W1-W2$, $g-z$ and $g-r$) for Sample II, the performance of a classifier improve a lot from 93.22 per cent to 99.15 per cent with additional infrared information when distinguishing stars and quasars. As a result, the infrared information had better be considered as possible when targeting quasar candidates, especially obscured quasars not identified only by optical information.

\cite{Stern05} and \cite{Hickox07} pointed out that infrared information is efficient to discriminate quasars. \cite{Bovy12} and \cite{Dip15} showed the power of infrared information for targeting quasar candidates using Extreme deconvolution (XD). Our work further proves that infrared information is of great importance to single out quasar candidates by photometric data.

Comparing the best results by two-layer binary classifiers in Tables~4-5 with those by multiclass classifier in Table~6, no matter with optical information or both optical and infrared information, Precision and Recall of quasars and stars by two-layer binary classifiers are better than those by multiclass classifier; however for galaxies, Recall of multiclass classifier is superior to that of two-layer binary classifiers, while Precision of two-layer binary classifiers outperforms that of multiclass classifier. As a result, predicted results by two-layer binary classifiers are more reliable for quasars and stars; predicted results by two-layer binary classifier are more trustworthy for galaxies in terms of Precision.

Except XGBoost, there are various methods used for celestial object classification, such as random forest, support vector machine (SVM). Table~7 shows the performance of random forest for classification of stars and quasars with Sample~I and Sample~II. For sample~I, the better performance is obtained with 9-features (${\Delta}g$, ${\Delta}r$, ${\Delta}z$, $g-r$, $r-z$, $g-z$, $r$, $g$, $z$); only given Accuracy, the best input pattern is 9-features (Accuracy: 93.27 per cent). For Sample~II, the best performance is achieved with 15-features (Accuracy: 99.14 per cent); Precision is 98.82 per cent and 99.53 per cent respectively for stars and quasars; Recall is 99.62 per cent and 98.54 per cent separately for stars and quasars. Comparing the results in Table~7 with those in Table~5, it is found that the performance of XGBoost is comparative to that of random forest whether for Sample~I or Sample~II and the running time of XGBoost is shorter than that of random forest. As a result, XGBoost outperforms random forest for our case in terms of efficiency.

\begin{table*}
	\begin{center}
		\caption[]{The performance of binary classifier for stars and quasars by Random Forest. \label{tab:comp3}}
		\begin{tabular}{rccccccc}
			\hline
			&& &\multicolumn{2}{c}{STAR} &\multicolumn{2}{c}{QSO} & \\
			\hline\noalign{\smallskip}
			Input pattern &Sample  &  Accu.(\%)  &Prec.(\%)& Rec.(\%)&Prec.(\%)&Rec.(\%)&Time(s)\\
			\hline
			(${\Delta}g$, ${\Delta}r$, ${\Delta}z$, $g-r$)     &Sample~I&85.51&83.52&89.99&88.07&80.63&125\\
			(${\Delta}g$, ${\Delta}r$, ${\Delta}z$, $g-r$, $r-z$, $r$)     &Sample~I&93.17&92.31&\textbf{94.81}&\textbf{94.16}&91.39&120 \\
			(\textbf{9-features})&Sample~I      &\textbf{93.27}&\textbf{92.83}&94.39&93.77&\textbf{92.05}&180 \\
			($g-r$, $r-z$, $r$, $g$, $z$, $w1$, $w2$)      &Sample~II&98.77&98.24&99.55&99.44&97.81&80\\
			(\texttt{15-features})  &Sample~II&\textbf{99.14}&\textbf{98.82}&\textbf{99.62}&\textbf{99.53}&\textbf{98.54}&120 \\
			\hline
			\multicolumn{8}{l}{$^a$ \texttt{9-features} represents ${\Delta}g$, ${\Delta}r$, ${\Delta}z$, $g-r$, $r-z$, $g-z$, $r$, $g$, $z$.}\\
			\multicolumn{8}{l}{$^b$ \texttt{12-features} represents $z-W2$,$z-W1$, $g-z$, $W1-W2$, $g-r$, ${\Delta}z$, $r-z$, $r-W2$, $r$, $z$, ${\Delta}g$, $g-W1$.}\\
			\multicolumn{8}{l}{$^b$ \texttt{15-features} represents  $z-W2$,$W1-W2$,$g-z$,$g-r$,$z-W1$,$r-z$,${\Delta}z$,$r$,$r-W2$,$z$,${\Delta}g$,$g-W1$,${\Delta}r$,$W1$,$g-W2$.}\\
		\end{tabular}
	\end{center}
\end{table*}

\section{Application of Classifiers}

According to the above experimental results for the known samples, we construct six classifier models, the detailed information is shown in Table~8.
If the binary classification model is adopted, two binary classifiers are needed to finish classification of galaxies, stars and quasars, the first classifier separates galaxies from stars and quasars, the second discriminates quasars and stars. While using the multiclass classification model, it only needs one multiclass classifier to finish classification of galaxies, stars and quasars. For each classifier, the optimal patterns obtained from above experiments are applied for different samples.

\begin{table*}
	\begin{center}
		\caption[]{ The six classifiers of XGBoost constructed}
		\begin{tabular}{rlll}
			\hline
			Classifier     &objective &Input pattern &Class  \\
			\hline
			Classifier $1^{st}$     &binary:logistic &(${\Delta}g$, ${\Delta}r$, ${\Delta}z$, $g-r$, $g-z$, $r$, $r-z$, $g$) &(point and extended sources) \\
			Classifier $2^{nd}$    &binary:logistic &(${\Delta}g$, ${\Delta}r$, ${\Delta}z$, $g-r$, $r-z$, $g-z$, $r$, $g$, $z$) &(stars and quasars) \\
			Classifier $3^{rd}$     &multi:softmax   &(${\Delta}g$, ${\Delta}r$, ${\Delta}z$, $g-r$, $g-z$, $r$, $r-z$, $z$ ) &(galaxies, stars and quasars) \\
			Classifier $4^{th}$     &binary:logistic &(\texttt{Pattern I}) &(point and extended sources) \\
			Classifier $5^{th}$     &binary:logistic &(\texttt{Pattern II}) &(stars and quasars) \\
			Classifier $6^{th}$     &multi:softmax   &(\texttt{Pattern III}) &(galaxies, stars and quasars) \\
			\hline
			\multicolumn{4}{l}{$^a$ Extended sources represent galaxies while point sources for stars and quasars.}\\
			\multicolumn{4}{l}{$^b$ \texttt{Pattern I} represents ${\Delta}g$, ${\Delta}z$, $g-W1$, $W1-W2$, $z-W1$, ${\Delta}r$, $g-z$, $z-W2$, $g-r$,$r-z$,$W1$, $r$, $g$, $z$, $r-W2$.}\\
			\multicolumn{4}{l}{$^c$ \texttt{Pattern II} represents  $z-W2$,$W1-W2$,$g-z$,$g-r$,$z-W1$,$r-z$,${\Delta}z$,$r$,$r-W2$,$z$,${\Delta}g$,$g-W1$,${\Delta}r$,$W1$,$g-W2$.}\\
			\multicolumn{4}{l}{$^d$ \texttt{Pattern III} represents $z-W2$, ${\Delta}z$, $W1-W2$, ${\Delta}r$, $g-r$, $z-W1$, ${\Delta}g$, $g-z$, $r-W2$, $r-z$, $r$. }
		\end{tabular}
	\end{center}
\end{table*}

After constructing these classifiers, we use these classifiers to predict BASS DR3 sources. Figure~4 shows the classification workflow. The red rectangle boxes indicate data analysis and black parallelograms represent intermediate data or results.

\begin{figure*}
\centering
\includegraphics[height=22cm,width=18cm]{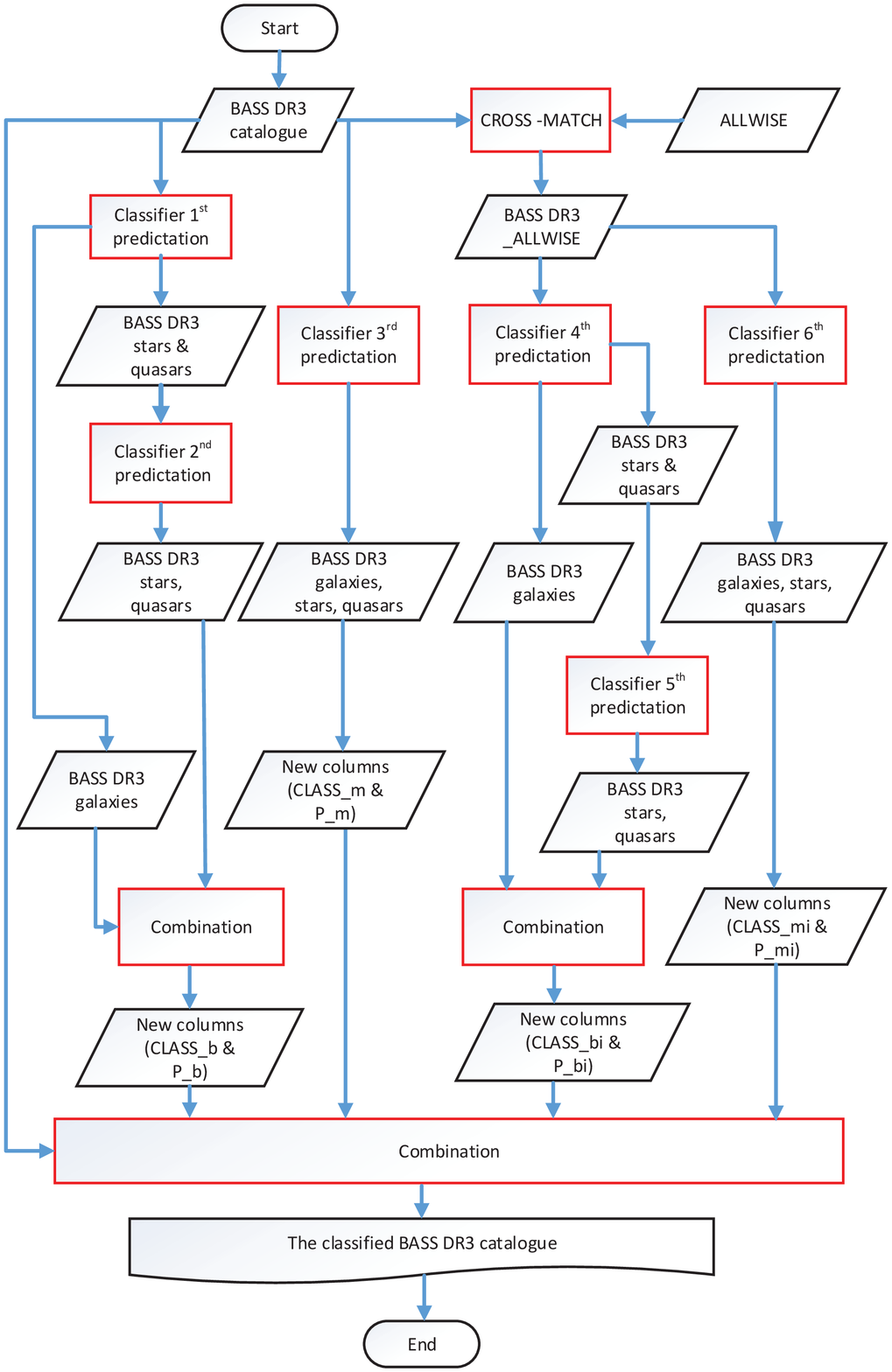}
\caption{The classification workflow.}
\label{fig5}
\end{figure*}


According to Figure~4, BASS-DR3 sources (110 896 598) are firstly classified into extended sources (galaxies) and point sources (stars and quasars) by Classifier~$1^{st}$, then the point sources are further separated into stars and quasars by Classifier~$2^{nd}$. BASS-DR3 sources are also directly divided into galaxies, stars and quasars by Classifier $3^{rd}$. By correlating BASS-DR3 sources with ALLWISE database, we obtain BASS-DR3-ALLWISE sources (43 859 467). Similar to classifying BASS DR3 sources, BASS-DR3-ALLWISE sources are firstly grouped into extended sources (galaxies) and point sources (stars and quasars) by Classifier~$4^{th}$, then the point sources are further discriminated into stars and quasars by Classifier~$5^{th}$. BASS-DR3-ALLWISE sources are directly distinguished into galaxies, stars and quasars by Classifier~$6^{th}$. Each prediction includes predicted classification label and likelihood. If the classification given by binary classifiers is the same as that by multiclass classifiers, the classification is more reliable. When classification is both provided by classifiers from optical information and classifiers from both optical and infrared information, the classification result is much trustier by classifiers from combined optical and infrared information. If only considering the results by binary classifier or multiclass classifier, the predicted results from two-layer binary classifier are adopted for quasars and stars; but for galaxies, the predicted results from two-layer binary classifiers are fit in terms of Precision while those form multiclass classifier are fit in terms of Recall.

For the BASS DR3 sources, all predicted results are combined in a whole table. The link address is http://paperdata.china-vo.org/Li.Changhua/bass/bassdr3-label-catalogue. Table~9 lists the 20 rows of predicted results, which is of great value for the further research on the characteristics and physics of BASS DR3 sources. The number of star, galaxy and quasar candidates by different classifiers with different information is shown in Table~10. When only using optical information, the BASS-DR3 sources simultaneously assigned as stars, galaxies and quasars by all classifiers respectively adds up to 19 829 533 ($P_{\rm S}>0.75$), 49 483 839 ($P_{\rm G}>0.75$) and 7 095 580 ($P_{\rm Q}>0.75$), among which quasar candidates with $P_{\rm Q}>0.90$ and $P_{\rm Q}>0.95$ amount to 2 775 970 and 1 166 517, respectively. When applying both optical and infrared information with the same predicted results by binary and multiclass classifiers, the number of star candidates is 12 785 232 ($P_{\rm S}>0.75$), 12 561 500 ($P_{\rm S}>0.90$) and 12 375 838 ($P_{\rm S}>0.95$); the number of galaxy candidates is 25 068 898 ($P_{\rm G}>0.75$), 21 890 547 ($P_{\rm G}>0.90$) and 18 606 073 ($P_{\rm G}>0.95$); the number of quasar candidates is 1 500 099 ($P_{\rm Q}>0.75$), 1 033 486 ($P_{\rm Q}>0.90$) and 798 928 ($P_{\rm Q}>0.95$). If only considering the completeness, we may combine the predicted results by binary and multiclass classifiers. Touching upon reliability, we adopt the same classified results by binary and multiclass classifiers only with optical information as those with both optical and infrared information. At this situation, the total number of quasar candidates is 1 262 964, among which there are 694 260 ($P_{\rm Q}>0.75$), 375 591 ($P_{\rm Q}>0.90$), 235 713 ($P_{\rm Q}>0.95$). As shown in Table 10 for larger than 95 per cent probability (i.e. both $P_{b1}$ and $P_{b2}$ above 95 per cent or both $P_{bi1}$ and $P_{bi2}$ above 95 per cent for two-layer binary classifiers, $P_{m}$ or $P_{mi}$ above 95 per cent for multiclass classifiers), the number of quasar candidates meanwhile by binary and multiclass classifiers is about 233 per deg$^2$ only with optical information, while the number of quasar candidates by binary, multiclass and binary\&multiclass classifiers is respectively about 218, 189 and 160 per deg$^2$ with optical and infrared information, which is consistent with the quasar luminosity function \citep{Pal16,Wilson19}. Based on optical and infrared information, the number of quasar candidates by binary classifier with different probabilities and those with larger than 95 per cent probability by different classifiers as function of $r$ magnitude are displayed in Figure~5 and the galactic location of the quasar candidates with larger than 95 per cent probability by binary\&multiclass classifiers is demonstrated in Figure~6. The number of quasar candidates decreases when $r>23$, which results from the number decrease of observed BASS sources as $r>23$. As shown in Figure~6, most of quasar candidates distribute on medium and high latitude, and there is no accumulation of quasar candidates along the galactic plane.

\begin{table*}
\begin{center}
	\tiny
\caption[]{Predicted results of BASS DR3 sources.}
 \begin{tabular}{lllcccccccccc}
 \hline
 id          &ra         & dec      &Class$\_b$  &$P_{b1}$  &$P_{b2}$ &Class$\_m$&$P_m$&Class$\_bi$&$P_{bi1}$&$P_{bi2}$&Class$\_mi$&$P_{mi}$\\
 \hline

95373011289 &135.04541765729505 &84.39853006107745 &0 &0.992 &0.993 &0 &0.981 &0 &0.997 &1.000 &0 &0.999 \\
95373012270 &137.06618290826282 &84.45571948189149 &0 &0.999 &0.965 &0 &0.994 &0 &0.999 &0.978 &0 &0.993 \\
95373014038 &135.24278336054908 &84.55902133199768 &0 &0.962 &0.999 &0 &0.963 &0 &1.000 &1.000 &0 &0.998 \\
95374009915 &145.63427167809962 &84.3344909394295 &0 &0.976 &0.999 &0 &0.980 &0 &0.956 &0.999 &0 &0.993 \\
95374010728 &145.64322900143938 &84.37462462607955 &0 &0.997 &0.993 &0 &0.956 &0 &1.000 &1.000 &0 &0.993 \\
95375006879 &146.35493428612986 &84.18153150948889 &0 &0.999 &0.996 &0 &0.983 &0 &1.000 &1.000 &0 &1.000 \\
95375008769 &146.7341471886791 &84.28327327388713 &0 &0.963 &0.980 &0 &0.970 &0 &0.999 &1.000 &0 &0.998 \\
95375010415 &146.783645308867 &84.37315314289253 &0 &0.999 &0.936 &0 &0.943 &0 &1.000 &1.000 &0 &0.999 \\
95371004745 &123.71207464080723 &84.1579130418392 &1 &1.000 &0.998 &1 &0.999 &1 &1.000 &1.000 &1 &1.000 \\
95371004752 &123.76995258941326 &84.15844883612738 &1 &0.989 &0.996 &1 &0.961 &1 &0.998 &0.999 &1 &0.997 \\
95371005233 &123.80162434791758 &84.18398863492929 &1 &0.999 &0.999 &1 &0.998 &1 &1.000 &1.000 &1 &1.000 \\
95371005325 &123.79929150615568 &84.18879820615973 &1 &0.996 &0.938 &1 &0.963 &1 &0.999 &0.999 &1 &0.999 \\
95371005366 &123.70596853052086 &84.19100324818724 &1 &0.970 &0.997 &1 &0.970 &1 &0.972 &0.997 &1 &0.998 \\
95371005498 &123.91422908428144 &84.19767799033559 &2 &0.965 & &2 &0.952 &2 &0.946 & &2 &0.968 \\
95371005594 &123.61974576953872 &84.2038370079812 &2 &0.960 & &2 &0.974 &2 &0.991 & &2 &0.991 \\
95371005759 &123.91886392365679 &84.21438506899153 &2 &0.995 & &2 &0.995 &2 &0.951 & &2 &0.990 \\
95371005789 &123.15220933495931 &84.21426532956772 &2 &0.986 & &2 &0.978 &2 &0.995 & &2 &0.974 \\
95371005812 &123.77907625197672 &84.21706698284555 &2 &0.995 & &2 &0.991 &2 &0.996 & &2 &0.983 \\
95371005951 &123.84673982200142 &84.22579958818966 &2 &0.973 & &2 &0.988 &2 &0.998 & &2 &0.988 \\
95371005960 &123.77794020713335 &84.22598315903795 &2 &0.912 & &2 &0.970 &2 &0.994 & &2 &0.991 \\
\hline
\multicolumn{11}{l}{$^a$ Class\_b is the classification label, $P_{b1}$ and $P_{b2}$ are respectively their classification probabilities by Classifier $1^{st}$  and Classifier $2^{nd}$;}\\
\multicolumn{11}{l}{$^b$ Class\_m is the classification label, $P_{m}$ is their classification probabilities by Classifier $3^{rd}$ ;}\\
\multicolumn{11}{l}{$^c$ Class\_bi is the classification label, $P_{bi1}$ and $P_{bi2}$ are respectively their classification probabilities by Classifier $4^{th}$ and Classifier $5^{th}$ ; }\\
\multicolumn{11}{l}{$^d$ Class\_mi is the classification label, $P_{mi}$ is their classification probabilities by Classifier $6^{th}$.}\\
\multicolumn{11}{l}{$^e$ The classification label definition: 0 represents quasars, 1 represents stars and 2 represents galaxies.} \\
\multicolumn{11}{l}{$^f$ Note for the sources assigned as galaxies, $P_{b2}$ or $P_{bi2}$ is default.} \\
\multicolumn{11}{l}{$^g$ This catalogue files is available in http://paperdata.china-vo.org/Li.Changhua/bass/bassdr3-label-catalogue. }\\
\multicolumn{11}{l}{$^h$ A portion is shown here for guidance regarding its form and content.}
\end{tabular}
\end{center}
\end{table*}

\begin{table*}
\begin{center}
\caption{Star, galaxy and quasar candidates by different classifiers with different information.}
\label{tab:anysymbols}
\begin{tabular}{lcccccc}	
\hline
Information &  & Optical && & Optical\&infrared & \\
\hline
Classifier  &  Binary & Multiclass & Binary\&Multiclass & Binary &Multiclass & Binary\&Multiclass \\
\hline
$P_{\rm S}>0.75$ &21 175 837 &20 550 441 &19 829 533 &12 938 789 &12 913 835 &12 785 232\\
$P_{\rm S}>0.90$ &18 570 067 &17 759 188 &17 043 293 &12 697 599 &12 711 753 &12 561 500\\
$P_{\rm S}>0.95$ &16 706 080 &15 727 548 &15 022 399 &12 519 421 &12 560 931 &12 375 838\\
\hline
$P_{\rm G}>0.75$ &54 360 301 &56 235 525 &49 483 839 &25 929 229 &26 490 981 &25 068 898  \\
$P_{\rm G}>0.90$ &41 243 840 &41 713 806 &35 544 793 &23 417 967 &23 899 117 &21 890 547 \\
$P_{\rm G}>0.95$ &32 053 408 &31 025 081 &25 949 348 &20 888 403 &21 088 855 &18 606 073 \\
\hline
$P_{\rm Q}>0.75$ &11 575 871 &9 459 579 &7 095 580 &2 078 981 &1 777 386 &1 500 099\\
$P_{\rm Q}>0.90$ & 5 271 397 &4 052 081 &2 775 970 &1 419 024 &1 221 362 &1 033 486\\
$P_{\rm Q}>0.95$ & 2 674 704 &1 874 389 &1 166 517 &1 088 976 &  943 486 & 798 928\\
\hline
\multicolumn{7}{l}{$^a$ $P_{\rm S}$ is the probabilities that sources are identified as stars, e.g. $P_{b1}$ and $P_{b2}$ are meanwhile above 95 percent if $P_{\rm S}>0.95$}\\
\multicolumn{7}{l}{\,\,\,\, for optical sample, $P_{bi1}$ and $P_{bi2}$ are meanwhile above 95 percent if $P_{\rm S}>0.95$ for optical and infrared sample;}\\
\multicolumn{7}{l}{$^b$ $P_{\rm G}$ is the probabilities that sources are identified as galaxies, e.g. $P_{b1}$ is above 95 percent if $P_{\rm G}>0.95$ for optical sample,}\\
\multicolumn{7}{l}{\,\,\,\, $P_{bi1}$ is above 95 percent if $P_{\rm G}>0.95$ for optical and infrared sample;}\\
\multicolumn{7}{l}{$^c$ $P_{\rm Q}$ is the probabilities that sources are identified as quasars, e.g. $P_{b1}$ and $P_{b2}$ are meanwhile above 95 percent if $P_{\rm Q}>0.95$}\\
\multicolumn{7}{l}{\,\,\,\, for optical sample, $P_{bi1}$ and $P_{bi2}$ are meanwhile above 95 percent if $P_{\rm Q}>0.95$ for optical and infrared sample.}\\
\end{tabular}
\end{center}
\end{table*}

\begin{figure*}
	\centering
    \includegraphics[bb=84 239 531 558,width=8cm,height=6.5cm]{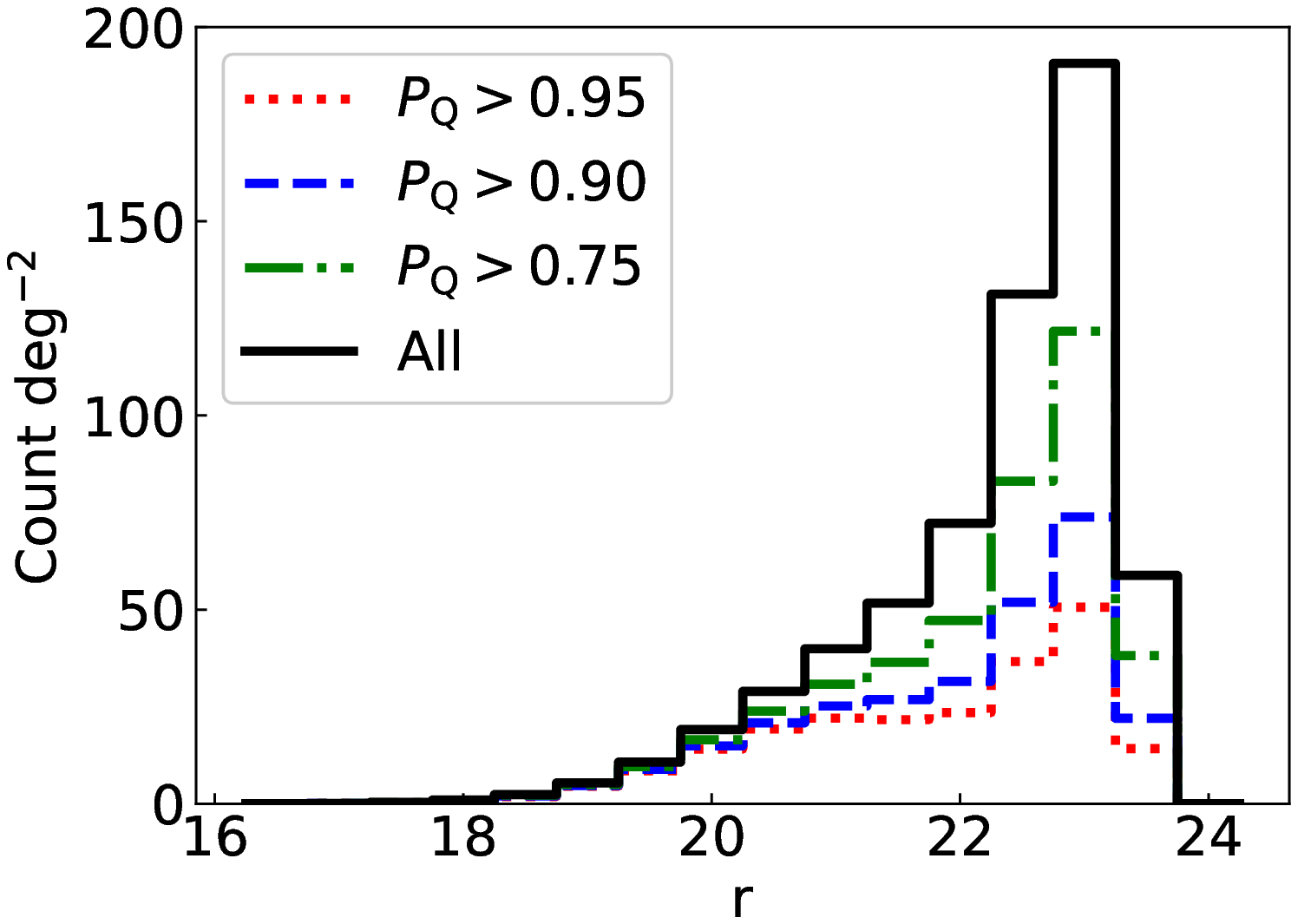}
	\includegraphics[bb=84 239 531 558,width=8cm,height=6.5cm]{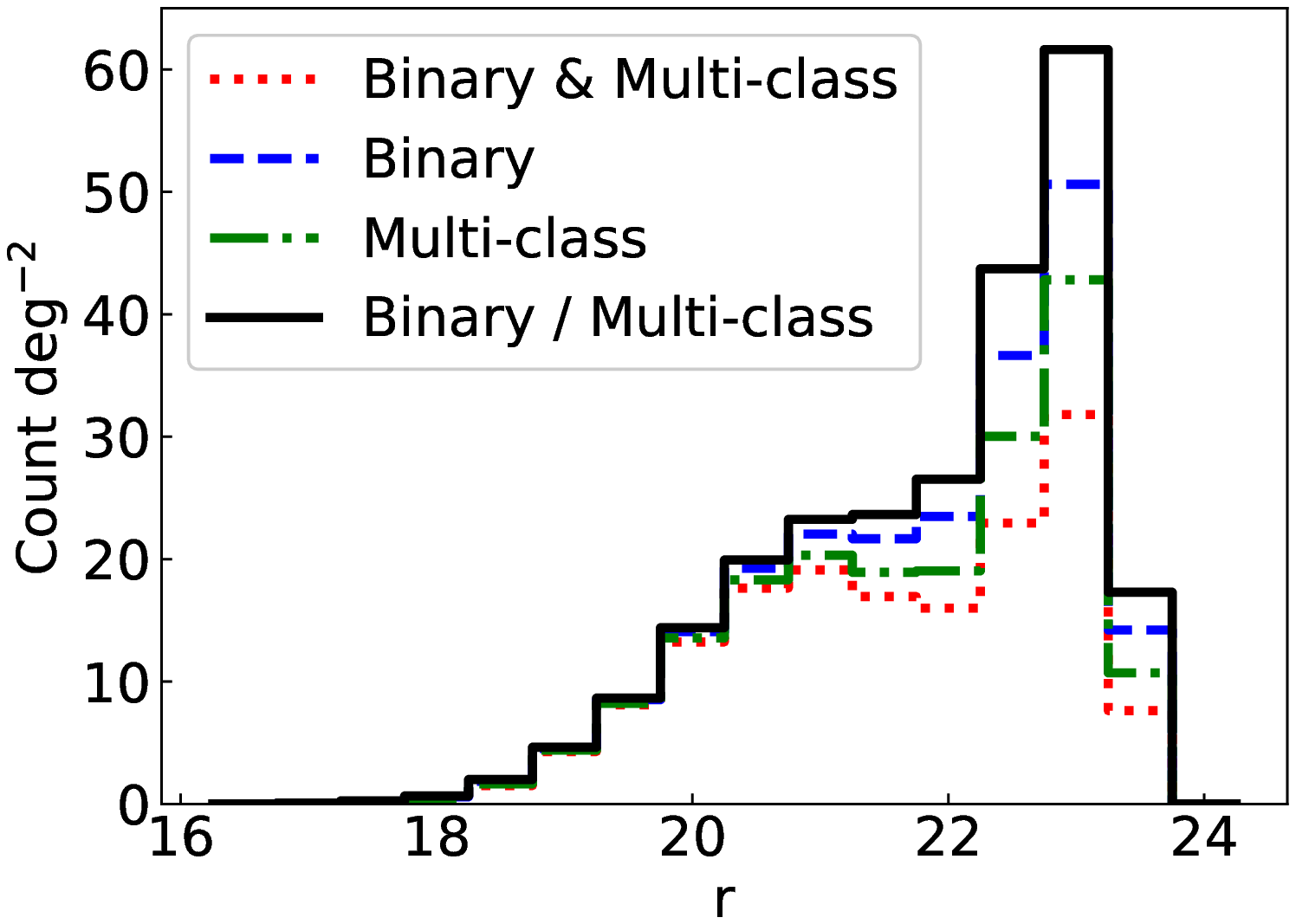}
	\caption{Left panel: the number of quasar candidates by binary classifier with optical and infrared information as function of $r$ magnitude for different probabilities ($P_Q>0.95 $ (red dotted line), $P_Q>0.90 $ (blue dash line), $P_Q>0.75 $ (green dotted dash line, all (black line)); right panel: the number of quasar candidates as function of $r$ magnitude with larger than 95 per cent probability by binary\&multiclass (red dotted line), binary (blue dash line), multiclass (green dotted dash line) and binary/multiclass (black line) classifiers based on optical and infrared information.}	
	\label{fig5}
\end{figure*}

\begin{figure*}
	\centering
	\includegraphics[bb=124 259 490 520,width=12cm,height=8cm]{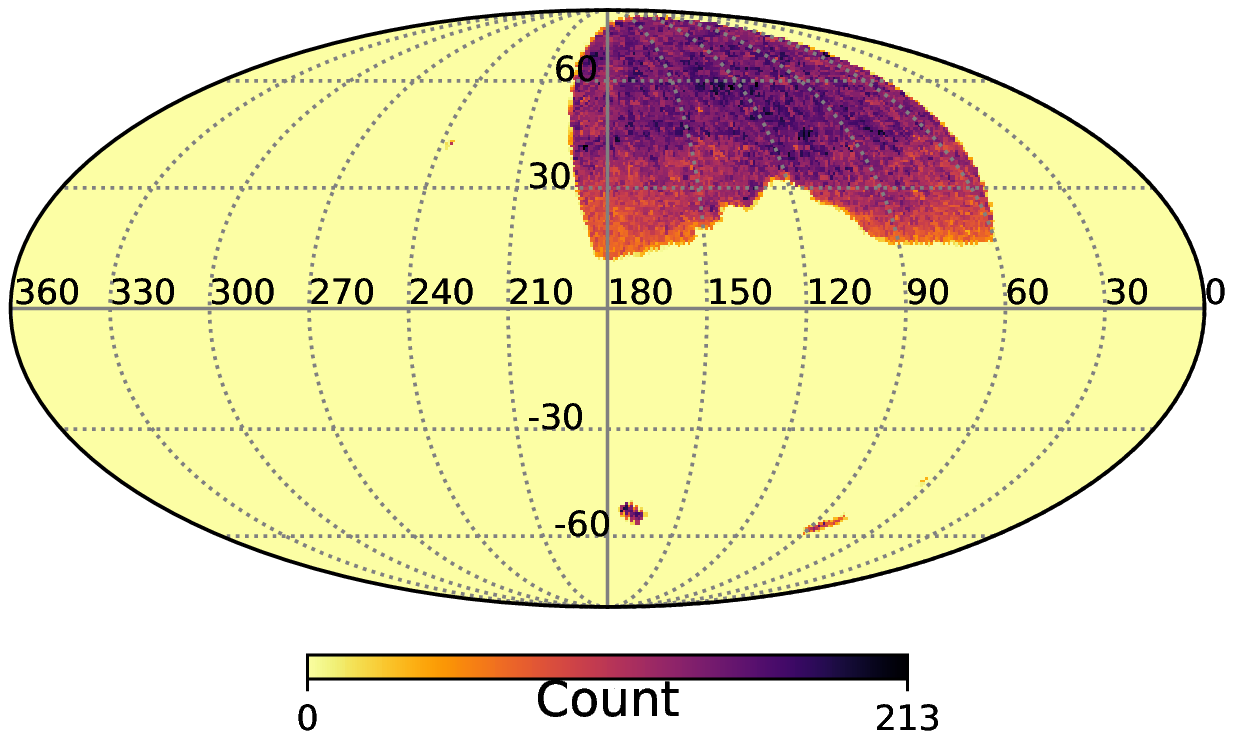}
 	\caption{The galactic location of quasar candidates with larger than 95 per cent probability by binary\&multiclass classifiers based on optical and infrared information.}	
	\label{fig6}
\end{figure*}

\section{Conclusions} \label{sec:conclusions}
It is hard to discriminate stars, galaxies and quasars only depending on single feature or two of all features. Facing classification in a high dimensional space, machine learning is a good choice. Our experimental results show that XGBoost classifiers get more efficient than colour cut or colour-colour plot, and are comparable to random forest on the classification of celestial objects. As for classification of more than two classes, multiclass classifier or multi-layer binary classifiers may perform; for our case, three-class classifier and two-layer binary classifiers are applied. We construct six classifiers to predict classification label and their probabilities with different input patterns for BASS-DR3 sources. The predicted results from different classifiers and original properties are listed in a whole table, which may shed light on further research about source properties of various kinds of objects in detail. The sources labelled as quasars will be taken as input catalogue of LAMOST, DESI or other projects for follow-up observation. With the future implementation of BASS survey, we may predict the new sources from the new survey by our classifiers or new classifiers created by more known spectroscopic objects. If only interested in some special objects, we may use machine learning to create the classifier of special objects and choose special object candidates for further study.

\section{Acknowledgements}
We are very grateful to the referees for their constructive suggestions to help us improve our paper. This work is supported by National Natural Science Foundation of China (NSFC)(11573019, 11803055, 11873066), the Joint Research Fund in Astronomy (U1531246, U1731125, U1731243, U1731109) under cooperative agreement between the NSFC and Chinese Academy of Sciences (CAS), the 13th Five-year Informatization Plan of Chinese Academy of Sciences (No. XXH13503-03-107). We would like to thank the National R\&D Infrastructure and Facility Development Program of China, ``Earth System Science Data Sharing Platform" and ``Fundamental Science Data Sharing Platform" (DKA2017-12-02-07). Data resources are supported by Chinese Astronomical Data Center (NADC) and Chinese Virtual Observatory (China-VO). This work is supported by Astronomical Big Data Joint Research Center, co-founded by National Astronomical Observatories, Chinese Academy of Sciences and Alibaba Cloud. This research has made use of BASS DR3 catalogue. BASS is a collaborative program between the National Astronomical Observatories of the Chinese Academy of Science and Steward Observatory of the University of Arizona. It is a key project of the Telescope Access Program (TAP), which has been funded by the National Astronomical Observatories of China, the Chinese Academy of Sciences (the Strategic Priority Research Program. The Emergence of Cosmological Structures grant no. XDB09000000), and the Special Fund for Astronomy from the Ministry of Finance. BASS is also supported by the External Cooperation Program of the Chinese Academy of Sciences (grant No. 114A11KYSB20160057) and the Chinese National Natural Science Foundation (grant No. 11433005). The BASS data release is based on the Chinese Virtual Observatory (China-VO).
The Guoshoujing Telescope (the Large Sky Area Multi-object Fiber Spectroscopic Telescope, LAMOST) is a National Major Scientific Project built by the Chinese Academy of Sciences. Funding for the project has been provided by the National Development and Reform Commission. LAMOST is operated and managed by the National Astronomical Observatories, Chinese Academy of Sciences.

We acknowledgment SDSS databases. Funding for the Sloan Digital Sky Survey IV has been provided by the Alfred P. Sloan Foundation, the U.S. Department of Energy Office of Science, and the Participating Institutions. SDSS-IV acknowledges support and resources from the Center for High-Performance Computing at the University of Utah. The SDSS web site is www.sdss.org. SDSS-IV is managed by the Astrophysical Research Consortium for the Participating Institutions of the SDSS Collaboration including the Brazilian Participation Group, the Carnegie Institution for Science, Carnegie Mellon University, the Chilean Participation Group, the French Participation Group, Harvard-Smithsonian Center for Astrophysics, Instituto de Astrof\'isica de Canarias, The Johns Hopkins University, Kavli Institute for the Physics and Mathematics of the Universe (IPMU) /University of Tokyo, Lawrence Berkeley National Laboratory, Leibniz Institut f\"ur Astrophysik Potsdam (AIP), Max-Planck-Institut f\"ur Astronomie (MPIA Heidelberg), Max-Planck-Institut f\"ur Astrophysik (MPA Garching), Max-Planck-Institut f\"ur Extraterrestrische Physik (MPE), National Astronomical Observatories of China, New Mexico State University, New York University, University of Notre Dame, Observat\'ario Nacional / MCTI, The Ohio State University, Pennsylvania State University, Shanghai Astronomical Observatory, United Kingdom Participation Group, Universidad Nacional Aut\'onoma de M\'exico, University of Arizona, University of Colorado Boulder, University of Oxford, University of Portsmouth, University of Utah, University of Virginia, University of Washington, University of Wisconsin, Vanderbilt University, and Yale University.

\section{Data availability}
The predicted results for BASS-DR3 sources is saved in a repository and can be obtained by a unique identifier, part of which is indicated in Table~9. It is put in paperdata at http://paperdata.china-vo.org, and can be available with http://paperdata.china-vo.org/Li.Changhua/bass/bassdr3-label-catalogue.

\end{document}